\documentclass[11pt,a4paper]{article}
\usepackage[utf8]{inputenc}
\usepackage[english]{babel}
\usepackage{amsmath}
\usepackage{amsthm}
\numberwithin{equation}{section}
\usepackage{amsfonts}
\usepackage{mathtools}
\usepackage{mathrsfs}
\usepackage{mathpazo}
\usepackage{multirow}
\usepackage{amssymb}
\usepackage{graphicx}
\usepackage{ifpdf}
\usepackage{braket}
\newcommand{\changefont}[3]{
\fontfamily{#1} \fontseries{#2} \fontshape{#3} \selectfont}
\changefont{ppl}{m}{n}
\usepackage[dvipsnames]{xcolor}

\newcommand{\be}{\mathbf{e}}
\newcommand{\bE}{\mathbf{E}}
\newcommand{\bF}{\mathbf{F}}
\newcommand{\bOmega}{\mathbf{\Omega}}
\newcommand{\barf}{\overline{f}}
\newcommand{\PB}[2]{\left\lbrace #1 , #2 \right\rbrace}

\usepackage{a4wide}

\usepackage[bookmarks=true,colorlinks=true,linkcolor=black,citecolor=BlueViolet,urlcolor=BlueViolet,bookmarksnumbered]{hyperref}
\usepackage{cite}

\begin{document}

\setcounter{equation}{0}
\setcounter{footnote}{0}
\setcounter{section}{0}

\thispagestyle{empty}

\begin{flushright} \texttt{LMU-ASC 31/19 \\ MPP-2019-197}\end{flushright}

\begin{center}
\vspace{1.5truecm}

{\Large \bf On current algebras, generalised fluxes and non-geometry}

\vspace{1.5truecm}

{David Osten}

\vspace{1.0truecm}

{\em Max-Planck-Institut f\"ur Physik \\
F\"ohringer Ring 6, 80805 M\"unchen, Germany}

\vspace{0.4truecm}
{\em Arnold Sommerfeld Center for Theoretical Physics\\
Theresienstraße 37, 80333 M\"unchen, Germany}

\vspace{1.0truecm}

{{\tt ostend@mpp.mpg.de}}

\vspace{1.0truecm}
\end{center}

\begin{abstract}
A Hamiltonian formulation of the classical world-sheet theory in a generic, geometric or non-geometric, NSNS background is proposed. The essence of this formulation is a deformed current algebra, which is solely characterised by the generalised fluxes describing such a background. The construction extends to backgrounds for which there is no Lagrangian description -- namely magnetically charged backgrounds or those violating the strong constraint of double field theory -- at the cost of violating the Jacobi identity of the current algebra.

The known non-commutative and non-associative interpretation of non-geometric flux backgrounds is reproduced by means of the deformed current algebra. Furthermore, the provided framework is used to suggest a generalisation of Poisson-Lie $T$-duality to generic models with constant generalised fluxes. As a side note, the relation between Lie and Courant algebroid structures of the string current algebra is clarified.
\end{abstract}

\pagebreak
\thispagestyle{empty}
\tableofcontents

\section{Introduction}
Non-linear $\sigma$-models \cite{GellMann:1960np} have been of great importance in particle physics and gravity. In particular two dimensional ones play an important role in the study of integrable models and string theory. In the latter they are descriptions of a string in a curved target space. The Lagrangian of such a $\sigma$-model is characterised by a metric $G$ and a 2-form gauge field $B$ on the target space.\footnote{We are be interested in the classical properties of the model and ignored the term proportional to the world-sheet Ricci scalar containing the dilaton, as it is not relevant to the classical (world-sheet) theory.} Studying the physical properties of these models reveals geometrical structure. Two prominent examples are the equations of motion and the 1-loop $\beta$-functions. Former is given by the geodesic equation to a torsionful connection -- the sum of the Levi-Cevita connection to the metric $G$ and a torsion term determined by the $\mathbf{H}$-flux, $\mathbf{H} = \mathrm{d}B$. The 1-loop $\beta$-function to the coupling on the other hand is given by the (generalised) Ricci tensor to this torsionful connection \cite{Brezin:1975sq,Friedan:1980jm,Braaten:1985is}. 

As long as the background is globally geometric -- meaning only diffeomorphisms and $B$-field gauge transformations are necessary for gluing coordinate patches -- metric and $B$-field are globally well-defined and seem to be an appropriate description of the background. But not all backgrounds in string theory can be described as such. So called non-geometric backgrounds have been shown to arise naturally as $T$-duals of geometric backgrounds \cite{Shelton:2005cf}. The ones we consider here can be understood as $T$-folds \cite{Hull:2004in,Hull:2006qs,Hull:2006va}, meaning that we allow for patching with $T$-duality transformations as well. They are expected to make up a big part of the landscape of string theory \cite{Flournoy:2004vn,Shelton:2006fd,Becker:2006ks,Font:2008vd,Plauschinn:2018wbo}, this includes not only duals of geometric backgrounds but also genuinely non-geometric backgrounds. These backgrounds can be described in terms of generalised geometry \cite{Hitchin:2004ut,Gualtieri:2003dx,Grana:2008yw} or the \textit{generalised fluxes}. These fluxes arise as parameters in gauged supergravities \cite{Aldazabal:2011nj,Geissbuhler:2011mx}, are the basis of a formulation of double field theory \cite{Tseytlin:1990va,Tseytlin:1990nb,Siegel:1993th,Siegel:1993bj,Hull:2009mi,Zwiebach:2011rg,Aldazabal:2013sca,Hohm:2013bwa,Geissbuhler:2013uka,Plauschinn:2018wbo} and have been shown to be related to the non-commutative and non-associative interpretations of these backgrounds \cite{Blumenhagen:2010hj,Lust:2010iy,Blumenhagen:2011ph,Condeescu:2012sp,Andriot:2012an,Chatzistavrakidis:2012qj,Mylonas:2012pg,Bakas:2013jwa}.

The aim of this article is to present a convenient formulation of the world-sheet theory which highlights the role of these generalised fluxes, making the non-geometric features more apparent than the not generally globally defined Lagrangian data $G$ and $B$. The key result of this article is that a \textit{Hamiltonian} description in terms of non-canonical coordinates on the string phase space achieves this objective. All the physical information about the background is encoded in a deformation of the Poisson structure
\begin{equation}
\Pi^\text{def} = \Pi^\eta + \Pi^\text{bdy.} + \Pi^\text{flux}.
\end{equation}
The canonical Poisson structure consists of an O$(d,d)$-invariant part $\Pi^\eta$ and a boundary contribution $\Pi^\text{bdy.}$, relevant for open strings and winding along compact directions. $\Pi^\text{flux}$ is characterised exactly by the generalised fluxes. Apart from $\Pi^{bdy.}$ this perspective already appeared back in \cite{Siegel:1993th,Siegel:1993bj} or in \cite{Alekseev:2004np} for geometrical $\mathbf{H}$-flux backgrounds. On the other hand non-geometric fluxes were already introduced as generalised WZ-terms in first order Lagrangians \cite{Halmagyi:2008dr,Halmagyi:2009te}, but only for a certain choice of generalised vielbein. Other perspectives on the connection of $\sigma$-models, current algebras and generalised geometry include \cite{Zucchini:2004ta,Zucchini:2005rh,Guttenberg:2006zi,Zabzine:2006uz,Belov:2007qj,Hekmati:2012fb}. In particular O$(d,d)$-invariant Hamiltonian setups and their non-geometric interpretation have been studied already in \cite{Blair:2013noa,Blair:2014kla}.

Before we state the main results and outline the structure of the paper, let us motivate the approach to our paper in two ways. The first one is a review of the Hamiltonian description of an electrically charged point particle in a magnetically charged background in electromagnetism. The second point is a collection of examples from the integrable models literature. Both points share the feature of a possible description by a free Hamiltonian and a deformed Poisson resp. symplectic structure.

\subsection{Point particle in an electromagnetic background}
\label{chap:EM}
As a motivational example, that shares many features with the string in NSNS backgrounds, let us consider a relativistic point particle with mass $m$ and electric charge $q$ in an arbitrary electromagnetic field \cite{Jackiw:1984rd,Jackiw:1985hq}. At first we define it by an electric potential $\mathbf{A} = A_\mu \mathrm{d} x^\mu$ with field strength $\mathbf{F}=\mathrm{d}\mathbf{A} = F_{\mu \nu} \mathrm{d}x^\mu \wedge \mathrm{d}x^\nu$. A convenient choice\footnote{The free Hamiltonian $H_{free} = \frac{e}{2m} \mathbf{p}^2$ with 4-momentum $\mathbf{p}$ is obtained via a Polyakov trick with the einbein $e$ so that $H_{free}$ is indeed the constraint corresponding to time reparameterisation invariance in this case. After gauge choice $e=1$ and minimal substitution we are left with above Hamiltonian.} of Hamiltonian, $H = \frac{1}{2m} \left( p - q A \right)^2$, together with the canonical Poisson structure gives the equations of motion
\begin{equation}
\dot{x}^\mu = \frac{1}{m} \pi^\mu \equiv \frac{1}{m} (p^\mu - q A^\mu) \quad \text{and} \quad \dot{\pi}_\mu = \frac{q}{m} F_{\mu \nu} \pi^\nu. \label{eq:DefinitionKinematicalMomentum}
\end{equation}
Alternatively this problem can be phrased in terms of new coordinates on the phase space $(x^i,\pi_i)$ with the kinematic momentum $\pi_\mu$. Let us note a few important characteristics of this formulation, which will also be key points in the string discussion:
\begin{itemize}
\item \textit{Preferred non-canonical phase space coordinates}. In terms of kinematic momentum $\pi^\mu$ the Hamiltonian is $H = \frac{\pi_\mu \pi^\mu}{2m}$, so we have a 'free' Hamiltonian. All background data - the coupling to the electromagnetic field - is encoded in the deformed Poisson brackets
\begin{equation}
\{x^\mu,x^\nu \} = 0,\quad \{ x^\mu , \pi_\nu \} = \delta^\mu_\nu, \quad \{ \pi_\mu , \pi_\nu \} = q F_{\mu \nu}, \label{eq:EMDeformedPB}
\end{equation}
resp. a the deformed symplectic structure $\omega = \omega_0 + q \mathbf{F}$. The Jacobi identity of the Poisson bracket resp. the closedness of $\omega$ is equivalent to the Bianchi identity in the standard Maxwell equations:
\begin{equation}
\mathrm{d} \omega = 0 \quad \Leftrightarrow \quad \mathrm{d} \mathbf{F}=0.
\end{equation}
The field equations for $\mathbf{F}$ can also be phrased conveniently in terms of the symplectic structure:
$\partial^\mu \omega_{\mu \nu} = 4\pi \mathbf{j}^{(e)}_\nu$.

\item \textit{Generalisation to magnetically charged backgrounds}. In this formulation there is no need to refer to the potential $A$, it is phrased only in terms of the field strength $F$. So it is well suited for generalisations to magnetically charged backgrounds with $\star \mathrm{d}\mathbf{F} = 4\pi j^{(m)}$.

Alternatively one could take another point of view, namely to consider this as a free particle in non-commutative or, in case $\mathrm{d} \mathbf{F} \neq 0$, even non-associative momentum space. This fact is basis for a example and toy model for the treatment of non-associative phase spaces \cite{Nambu:1973qe,Takhtajan:1993vr,DeBellis:2010pf,Mylonas:2013jha,Bojowald:2014oea}. Recently it has been shown that such a non-associative, or almost symplectic, phase space can be realised in a higher dimensional symplectic one \cite{Kupriyanov:2018xji,Marmo:2019ulx}.

\item \textit{Charge algebra.} This coordinate change in phase space (a symplectomorphism in the case without magnetic sources\footnote{To make this problem symmetric in electric and magnetic terms we could consider a dyon $(q,g)$ in an electromagnetic background $F$, e.g. a particle with Lorentz force
$\dot{\pi}_\mu = \frac{1}{m} \left( q F_{\mu \nu} + g \tilde{F}_{\mu \nu} \right) \pi^v$, thus corresponding to the deformed symplectic structure is $\omega = \omega_0 + q \mathbf{F} + g \star \mathbf{F}$, which is not symplectic anymore, as soon as we have any electric or magnetic sources for $F$. For the dyon then there is no (local) field redefinition anymore connecting the two formulations.}) is simply the local field redefinition from canonical to kinematic momenta
\begin{equation}
\omega = - \mathrm{d} \theta = \mathrm{d} (p_\mu \mathrm{d} x^\mu) = \mathrm{d} (\pi_\mu + q A_\mu) \wedge \mathrm{d} x^\mu = \mathrm{d}\pi_\mu \wedge \mathrm{d}x^\mu + q \mathbf{F}. \label{eq:SymplecticStructureEMtwisted}
\end{equation}
\end{itemize}

\subsection{Integrable models and deformations of current algebras} 
\label{chap:IntModel}
The principal chiral model, the theory of the embedding of a classical string into a group manifold $G$, is one of the most important toy models for the study of integrable $\sigma$-models. It can be defined by a Hamiltonian
\begin{equation}
H = \frac{1}{2} \int \mathrm{d}\sigma  \left(\kappa^{ab} j_{0,a} j_{0,b} + \kappa_{ab} j_1^a j_1^b \right) \label{eq:PCMHamiltonian}
\end{equation}
and the following Poisson structure, the current algebra,
\begin{align}
\PB{j_{0,a}(\sigma)}{j_{0,b}(\sigma^\prime)} &= - {f^c}_{ab} j_{0,c}(\sigma) \delta(\sigma - \sigma^\prime) \nonumber \\
\PB{j_{0,a}(\sigma)}{j_{1}^b(\sigma^\prime)} &= - {f^b}_{ca} j_1^c (\sigma) \delta(\sigma - \sigma^\prime) - \delta^a_b \partial_{\sigma^\prime} \delta(\sigma - \sigma^\prime) \label{eq:PCMCurrentAlgebra} \\
\PB{j_{1}^a(\sigma)}{j_{1}^b(\sigma^\prime)} &=  0. \nonumber
\end{align}
${f^c}_{ab}$ are structure to the Lie algebra $\mathfrak{g}$ of $G$ and $\kappa$ its Killing form. We take $\partial_\tau = \PB{\cdot}{H}$. The Hamiltonian equations of motion contain both the flatness condition $\mathrm{d}j + \frac{1}{2}[j,j]=0$ and the equations of motion $\mathrm{d}\star j =0$. We have $j_{0,a} = (g^{-1} \partial_0 g)_a = p_a$ and $j_1^a = (g^{-1} \partial_1 g)^a = {e_i}^a \partial x^i$. This identification will be different for distinct current algebras/models and is what we later call a (generalised) frame. We still have to define the brackets between the $j_{\alpha}$ and functions $f$ on $G$: \begin{small}
\begin{equation}
\PB{j_{0,a}(\sigma)}{f(x(\sigma^\prime))} = -\partial_a f(x(\sigma)) \delta(\sigma - \sigma^\prime) \equiv - {e_a}^i \partial_i f(x(\sigma)) \delta(\sigma - \sigma^\prime), \quad \PB{j_{1,a}(\sigma)}{f(x(\sigma^\prime))} = 0, \nonumber
\end{equation}
\end{small} where we chose some coordinates $x$ on $G$.

The principal chiral model possesses many deformations which preserve one its most interesting properties: its classical integrability. Although integrability will not be the main focus of this paper, one detail of these integrable deformations motivates our approach - the deformations can be understood as deformations of the current algebra \eqref{eq:PCMCurrentAlgebra} instead of the deformation of a Hamiltonian or Lagrangian.
\begin{itemize}
\item The introduction of a WZ-term in the Lagrangian can be accounted for by a change of the $j_0$-$j_0$ Poisson bracket in comparison to \eqref{eq:PCMCurrentAlgebra}
\begin{align}
\PB{j_{0,a}(\sigma)}{j_{0,b}(\sigma^\prime)}_{WZW} &= - \left( {f^c}_{ab} j_{0,c}(\sigma) + k f_{abc} j_1^c(\sigma) \right) \delta(\sigma - \sigma^\prime) \label{eq:WZWCurrentAlgebra}.
\end{align}
Classically $k$ can be considered as a deformation parameter. See for example the standard textbook \cite{Faddeev:1987ph} for more details on the Hamiltonian treatment of the WZW-model.

\item The $\sigma$-model Lagrangian of the $\eta$-deformation was discovered in \cite{Klimcik:2002zj,Klimcik:2008eq} and its target space interpretation as a $q$-deformation of the original group manifold was given in \cite{Delduc:2013fga}. It can also be represented by a modication of the current algebra, as such it arose already in \cite{Faddeev:1985qu}. Compared to \eqref{eq:PCMCurrentAlgebra} the Poisson bracket between the $j_1$ is:
\begin{align}
\PB{j_{1}^a(\sigma)}{j_{1}^b(\sigma^\prime)}_\eta &=  \frac{\eta^2}{1 - \eta^2}f^{abc} j_{0,c}(\sigma) \delta(\sigma - \sigma^\prime). \label{eq:etaCurrentAlgebra}
\end{align}

\item The $\lambda$-deformation was introduced directly in terms of a deformation of the current algebra, originally for $G=$SU$(2)$ in \cite{Balog:1993es} and later generalised to arbitrary groups in \cite{Sfetsos:2013wia}, accompanied with a Lagrangian derivation. It can be completed to supergravity solutions, corresponds to certain $q$-deformations of the original group and has been argued to be equivalent via Poisson-Lie $T$-duality and analytic continuation of the deformation parameter $\eta \leftrightarrow \pm i \lambda$ to the $\eta$-deformation \cite{Sfetsos:2014cea,Sfetsos:2015nya,Klimcik:2015gba,Borsato:2016ose}.

Again after some rescaling of the currents compared to the original articles the $\lambda$-deformation corresponds only to a change in the $j_1$-$j_1$-Poisson bracket:
\begin{align}
\PB{j_{1}^a(\sigma)}{j_{1}^b(\sigma^\prime)}_\lambda &= - \frac{\lambda^2}{1 + \lambda^2}f^{abc} j_{0,c}(\sigma) \delta(\sigma - \sigma^\prime). \label{eq:lambdaCurrentAlgebra}
\end{align}
Phrased like this in the Hamiltonian formalism and compared to \eqref{eq:etaCurrentAlgebra}, we see directly that $\lambda$- and $\eta$-deformations are equivalent via analytic continuation $\eta \leftrightarrow  \pm i \lambda$. 
\end{itemize}
With this short survey we have motivated that in the Hamilton formulation deformations of the current algebra are a convenient playground. In fact we will see that every bosonic string $\sigma$-model can be represented by the free Hamiltonian and a modified current algebra. 

A related discussion of the SU$(2)$ principal chiral model aimed on the features connected the generalised geometry can be found in \cite{Marotta:2019wfq}. 

\subsection{Main results and overview}

\paragraph{Strings in arbitrary $\sigma$-model backgrounds}

Let us define the Hamiltonian theory of a string in an NSNS background characterised by the (geometric and globally non-geometric) generalised fluxes $\bF_{ABC}$. In terms of phase space variables $\bE_A(\sigma)$ the Hamiltonian takes the form of a 'free' Hamiltonian
\begin{equation}
H = \frac{1}{2}\oint \mathrm{d}\sigma \ \delta^{AB} \bE_A(\sigma) \bE_B(\sigma),
\end{equation}
whereas the background data, namely the generalised fluxes $\bF_{ABC}$ is encoded in the deformed current algebra
\begin{align}
\PB{\mathbf{E}_A(\sigma_1)}{\mathbf{E}_B(\sigma_2)} &=  \frac{1}{2}\eta_{AB} (\partial_1- \partial_2) \delta(\sigma_1 - \sigma_2)  - {\bF^C}_{AB}(\sigma) \mathbf{E}_C(\sigma) \delta(\sigma_1 - \sigma_2) \nonumber  \\
&{} \quad + \text{boundary term}  , \label{eq:IntroPBDeformedCurrent}
\end{align}
where $\eta$ is the O$(d,d)$-metric. The $\bE_A$ are a priori abstract, but from a Lagrangian perspective they are connected to Darboux coordinates ($x^i$, $p_i$) via $\bE_A = {E_A}^I \bE_I = {E_A}^I  \left( p_i , \partial x^i \right)$. ${E_A}^I$ is a generalised vielbein. The Hamiltonian equations of motion of a string in a generic background take a convenient form. We recognise them as a Maurer-Cartan equation pulled back to the world-sheet.
\begin{equation}
d\mathcal{E}^A + \frac{1}{2} {\bF^A}_{BC}\mathcal{E}^B \wedge \mathcal{E}^C = 0 \quad \text{and} \quad  \mathcal{E}_A =  \delta_{AB} \star \mathcal{E}^B.
\end{equation}
The connection to a $\sigma$-model Lagrangian respectively a choice of 'Darboux coordinates' is given by a generalised vielbein ${E_A}^I$, s.t.
\begin{equation}
\bF_{ABC} = \left( \partial_{[A} {E_B}^I \right) E_{C]I}
\end{equation}
with $\partial_A = {E_AI}^I \partial_I$. Such a frame exists if the Jacobi identity of the current algebra \eqref{eq:IntroPBDeformedCurrent}, which is equivalent to the Bianchi identity of generalised fluxes
\begin{equation}
\partial_{[A} \mathbf{F}_{BCD]} - \frac{3}{4} {\mathbf{F}^E}_{[AB} \mathbf{F}_{CD]E} = 0, \label{eq:IntroBianchi}
\end{equation}
is fulfilled.

\paragraph{Brackets on the phase space}

Let us summarise the different local forms of canonical (Poisson) brackets, which we will discuss in that article, phrased in an O$(d,d)$-covariant way.
\begin{small}
\begin{align}
\text{current bracket}: \quad \PB{\bE_I(\sigma_1)}{\bE_J(\sigma_2)} &= \frac{1}{2} \eta_{IJ} (\partial_1 - \partial_2) \delta(\sigma_2 - \sigma_1) + \frac{1}{2}\omega_{IJ} \int \mathrm{d} \sigma \partial \big( \delta(\sigma - \sigma_1) \delta(\sigma - \sigma_2) \big) \nonumber \\
\text{Poisson bracket}: \quad \PB{x^i(\sigma_1)}{p_j(\sigma_2)} &= \delta^i_j \delta(\sigma_1 - \sigma_2) =0 \label{eq:IntroPBCanonicalCurrent} \\
\text{DFT bracket:} \quad \PB{X_I(\sigma_1)}{X_J(\sigma_2)} &= \eta_{IJ} \bar{\Theta}(\sigma_1 - \sigma_2) \quad \text{with} \quad  \ \bar{\Theta}(\sigma) = \frac{1}{2} \text{sign}(\sigma) \nonumber 
\end{align}\end{small}
The unspecified boundary term in \eqref{eq:IntroPBDeformedCurrent} stems from the second term in canonical current algebra. Here we have that
\begin{align*}
\bE_I(\sigma) = (p_i(\sigma), \partial x^i(\sigma)) \quad \text{and} \quad \omega = \left( \begin{array}{cc} 0 & - \mathbb{1} \\ \mathbb{1} & 0 \end{array} \right).
\end{align*}
For the open string it gives a boundary contribution but also for closed strings it might give a one, e.g. from winding along compact directions.  Of the three properties -- skew-symmetry, Jacobi identity and O$(d,d)$-invariance -- Lie, Courant and Dorfman bracket satisfy two each and the third one up to such a total derivative term under the $\sigma$-integral. E.g. the above form \eqref{eq:IntroPBCanonicalCurrent} is the Lie bracket on sections of $(T \oplus T^\star) LM$, where $M$ is the target space and $LM$ denotes the configuration space of the string in $M$. Without the second term \eqref{eq:IntroPBCanonicalCurrent} would be a Courant bracket, which is O$(d,d)$-invariant and skew-symmetric but violates the Jacobi identity by such a total derivative term under the $\sigma$-integral. 

This second term has been neglected in previous literature but becomes crucial for the non-geometric interpretation of the current algebra. For example only when it is considered the current algebra of the locally geometric pure $\mathbf{Q}$-flux background is associative as expected. This is shown in section \ref{chap:NonGeometry}.

\paragraph{Generalisations to magnetically charged and double field theory backgrounds} 
The world-sheet theory as a Lagrangian $\sigma$-model is only defined in 'electric' backgrounds. I.e. these are those that fulfil \eqref{eq:IntroBianchi}, and are locally geometric. 

Instead the Hamiltonian formulation in the generalised flux frame extend straightforwardly to magnetically charged and locally non-geometric backgrounds. If we have a magnetically charged background the Bianchi identity of generalised fluxes \eqref{eq:IntroBianchi} is not fulfilled. This means we cannot find a generalised vielbein that will connect the deformed current algebra \eqref{eq:IntroPBDeformedCurrent} to the canonical one \eqref{eq:IntroPBCanonicalCurrent}. Analogously to the case of the point particle in an magnetic monopole background the violation of the Bianchi identity corresponds to a violation of the Jacobi identity of the current algebra.

In order to study the world-sheet theory in a double field theory background we do not need to double the phase space. The dual field $\tilde{x}$ is not independent and given by $p_i(\sigma) = \partial \tilde{x}_i(\sigma)$. Allowing for a dependence of the generalised vielbein on the original as well as the dual coordinates might induce an additional strong constraint violating term in the deformed current algebra \eqref{eq:IntroPBDeformedCurrent}. This term will be non-local in general and leads to a modification of the Virasoro algebra of world-sheet diffeomorphisms. But, even if the background generalised vielbein fulfils the strong constraint by itself, the Jacobi identity of generic functions of this doubled phase space is only fulfilled up to strong constraint violating terms, e.g.
\begin{equation}
\PB{\Psi}{\PB{\phi_1}{\phi_2}} + c.p. = \int \mathrm{d} \sigma_1 \mathrm{d}\sigma_2 \ \frac{1}{2}\left(\eta_{KL} + \omega_{KL} \right) \phi_{[1}^K(\sigma_2) \frac{\delta \Psi}{\delta X^I(\sigma_1)} \frac{\delta \phi_{2]}^L(\sigma_2)}{\delta X_I(\sigma_1)} + \ \text{other terms,} \nonumber
\end{equation}
where $\phi_i = \int \mathrm{d} \phi_i^I(\sigma) \mathbf{E}_I(\sigma)$ and $\Psi$, $\phi_i^I(\sigma)$ are (multilocal) functionals of the fields $X_I(\sigma)= (\tilde{x_i}(\sigma),x^i(\sigma))$. See section \ref{chap:DFT} for more details.

\paragraph{Non-geometric interpretation}
Going to the generalised flux frame is very convenient in the current algebra. Given such a deformed current algebra, we decompose $\bE_A$ to $(e_{0,a}(\sigma) , e_1^a(\sigma) )$ and define $\partial y^a = e_1^a$. These coordinates $y^a$ are the ones which  show a potentially non-geometric, e.g. non-commutative or non-associative, behaviour. We obtain their Poisson brackets simply by integrating the deformed current algebra \eqref{eq:IntroPBCanonicalCurrent}.

This is in contrast to many previous derivations of the non-geometric nature of the backgrounds which relied on finding mode expansions first. In section \ref{chap:NonGeometry} we show that we reproduce the known results on open strings in a constant $B$-field background and closed strings in a constant $\mathbf{Q}$-flux background.

\paragraph{Classical generalised $T$-dualities} 
The framework easily realises abelian $T$-duality. For Poisson-Lie $T$-dualisable resp. the $\mathcal{E}$-models \cite{Klimcik:1995dy,Klimcik:2015gba} the current algebra is exactly of the kind \eqref{eq:IntroPBDeformedCurrent} with 
\begin{align*}
{\mathbf{F}^c}_{ab} = {f^c}_{ab}, \qquad {\mathbf{F}_c}^{ab} = {\barf_c}^{ab} \qquad \text{and} \qquad \mathbf{F}_{abc} = \mathbf{F}^{abc} = 0,
\end{align*}
where the constants ${f^c}_{ab}$ and ${\barf_c}^{ab}$ are structure constants to a Lie bialgebra \cite{Sfetsos:1997pi,Demulder:2018lmj}. The duality transformations are linearly realised in that basis. We show that for certain parameterisations of $\bF_{ABC}$ there exists an extension of Poisson-Lie $T$-duality, which we call \textit{Roytenberg duality}. It nevertheless relies on the same trick as Poisson-Lie $T$-duality, namely that a Poisson-bivector on a group manifold realises a constant generalised flux background. 

We discuss these (generalised) $T$-dualitiesthe as canonical transformation. The generating functions for these can be implicitly defined by $\mathcal{F}_{\mathfrak{Q}_{AB}} = - \eta^{AB} \mathfrak{Q}_{AB}$, where the $\mathfrak{Q}_{AB}$ are $\mathfrak{o}(d,d)$-charges on the phase space fulfilling $\PB{\mathfrak{Q}_{[AB]}}{\bE_C} = \eta_{C[A}\bE_{B]}$.

\paragraph{Structure of the paper}
Section \ref{chap:reviewGeneralisedFluxes} sets conventions and aims to clarify our interpretation of the generalised flux backgrounds. The review of the string in an $\mathbf{H}$-flux background in section \ref{chap:KalbRamond} sets the basis for further discussion and introduces the $\mathbf{H}$-flux as a twist of symplectic structure. Based on observations in that section we distinguish the algebraic structures of the phase space in section \ref{chap:CourantAlgebroid}. 

The central result -- the formulation of string theory in an arbitrary generalised flux background by a free Hamiltonian but a deformed current algebra -- is derived in section \ref{chap:Hamiltonian}.1. This includes a discussion on the general form of the equations of motion, the Virasoro constraints,  the boundary term and the generalisation to magnetically charged backgrounds.  In section \ref{chap:Hamiltonian}.2 and \ref{chap:DFT} the investigation on (generalised) $T$-dualities and motivated by this a brief extension of the previous discussion to double field theory backgrounds follow.

In section \ref{chap:NonGeometry} we propose a direct non-geometric interpretation of the deformed current algebra and confirm that it reproduces the standard results of an open string in a constant $B$-field backgrounds and of a closed string in a constant $\mathbf{Q}$-flux background. We close with some outlooks and potential further directions.

\section{Preliminaries}
\label{chap:review}

\subsection{Generalised fluxes and non-geometric backgrounds}
\label{chap:reviewGeneralisedFluxes}

In this section we collect and review well-known material about generalised geometry and generalised fluxes in order to clarify our conventions and prepare later discussions. We refer to standard reviews of double field theory \cite{Hull:2009mi,Zwiebach:2011rg,Aldazabal:2013sca,Hohm:2013bwa,Plauschinn:2018wbo}, generalised geometry \cite{Gualtieri:2003dx,Grana:2008yw} and the generalised flux formulation \cite{Geissbuhler:2013uka} for more details.

\paragraph{Generalised metric and generalised vielbeins}
We define O$(d,d)$-transformations to be $2d\times 2d$-matrices, which leave the O$(d,d)$-metric
\begin{equation}
\eta = \left( \begin{array}{cc} 0 & \mathbb{1} \\ \mathbb{1} & 0 \end{array} \right)
\end{equation}
invariant. The action of an $M \in$ O$(d,d)$ on a (bosonic) string background, characterised by a metric $G$ and a 2-form gauge field $B$, or equivalently by the generalised metric
\begin{equation}
\mathcal{H}(G,B) = \left(\begin{array}{cc} G - B G^{-1} B & B G^{-1} \\ - G^{-1} B & G^{-1} \end{array} \right), \label{eq:GeneralisedMetric}
\end{equation} 
is given by $ \mathcal{H}(G^\prime,B^\prime) = M \mathcal{H}(G,B) M^T$. A \textit{generalised vielbein} or \textit{frame} ${E_A}^I(x)$ is defined to be any (local) O$(d,d)$-transformation in the component connected to the identity that diagonalises and trivialises the generalised metric, i.e.
\begin{equation}
{E_A}^I {E_B}^J \eta_{IJ} = \eta_{AB} \quad \text{and} \quad  {E_A}^I {E_B}^J \mathcal{H}_{IJ}  = \gamma_{AB} := \left( \begin{array}{cc} \gamma_{ab} & 0 \\ 0 & \gamma^{ab} \end{array} \right), \label{eq:GeneralisedFluxFrame}
\end{equation}
where $\gamma$ is a flat metric in the signature of the target space and is used to raise and lower indices $a,b,... = 1,...,d$. Indices $A,B,... = 1,...,2d$ denote the 'flat' indices and are raised and lowered by $\eta_{AB}$. Unless stated otherwise we will assume that the generalised vielbeins are (local) functions on the original target space with coordinates $x^i$. With this assumption we restrict to \textit{locally geometric} backgrounds, but below and in section \ref{chap:DFT} we will also discuss the generalisation to \textit{locally non-geometric} backgrounds.

All generalised vielbeins can be generated by successively performing
\begin{align}
B\text{-shifts:} \ E^{(B)} &= \left( \begin{array}{cc} \mathbb{1} & B \\ 0 &\mathbb{1}\end{array} \right), \quad GL\text{-transformations} \ E^{(e)} = \left( \begin{array}{cc} e & 0 \\ 0 & (e^{-1})^T \end{array} \right) \label{eq:OddBasis}\\
\beta\text{-shifts:} \ E^{(\beta)} &= \left( \begin{array}{cc} \mathbb{1} & 0 \\ \beta &\mathbb{1}\end{array} \right), \quad \text{factorised dualities:} \ E^{(T_i)} = \left(\begin{array}{cc} \mathbb{1} - \delta_i & \delta^i \\ \delta_i &\mathbb{1} - \delta_i \end{array} \right) \nonumber
\end{align}
for skewsymmetric $d\times d$-matrices $B$ and $\beta$, an invertible matrix $e$ (a $d$-dimensional vielbein) and $(\delta_i)_{jk} = \delta_{ij} \delta_{ik}$.

\paragraph{Weitzenb\"ock connection and generalised fluxes}
The generalised Weitzenb\"ock connection of such a generalised flux frame is defined by
\begin{equation}
\bOmega_{C,AB} = \partial_C {E_A}^I {E_{BI}} \quad \text{with} \quad \partial_A := {E_A}^I \partial_I, \label{eq:WeitzenboeckCon}
\end{equation}
fulfilling ${\bOmega^C}_{AB} = - {\bOmega^C}_{BA}$ due to \eqref{eq:GeneralisedFluxFrame}. $\partial_I = (\partial_i,\partial^i)$, where $\partial^i$ denotes the derivative w.r.t. to dual coordinates $\tilde{x}_i$, which vanishes for locally geometric backgrounds. In fact only the totally skewsymmetric combination\footnote{We use the conventions: \begin{align*}
v_{[a}w_{b]} &= v_a w_b - v_b w_a, \qquad u_{[a} v_{b}w_{c]} =u_a v_b w_c + \ \text{\textit{cyclic} perm.} \\ 
u_{[a} v_{b}w_{c} z_{d]} &= u_{a} v_{b} w_c z_{d} + (-1)^{\text{sign} } \times \text{\textit{all} permutations}
\end{align*}} will be relevant for us: the \textit{generalised flux}
\begin{equation}
\bF_{ABC} := \bOmega_{[C,AB]} = \left( \partial_{[{A}} {E_{{B}}}^I \right) {E_{{C}]I}} . \label{eq:GeneralisedFlux}
\end{equation}
It includes the four fluxes -- $\mathbf{H}$, $\mathbf{f}$, $\mathbf{Q}$ and $\mathbf{R}$ -- for different choices of the indices on the $\bF_{ABC}$
\begin{align}
{\mathbf{H}}_{abc} &\equiv {\bF}_{abc}, \qquad {\mathbf{f}^c}_{ab} \equiv {\bF^c}_{ab} = {{\bF_b}^c}_a = {\bF_{ab}}^c \nonumber \\
\mathbf{R}^{abc} &\equiv {\bF}^{abc} ,\qquad {\mathbf{Q}_c}^{ab} \equiv {\bF_c}^{ab} = {{\bF^b}_c}^a = {\bF^{ab}}_c \nonumber
\end{align}
In a \textit{generalised flux frame}  \eqref{eq:GeneralisedFluxFrame} all the information about the background is stored inside the generalised fluxes, instead of the generalised metric. The generalised metric will be trivial in that frame.

\paragraph{Bianchi identities}
Generalised fluxes, given as above in terms of a generalised vielbein, cannot be chosen arbitrarily but have to fulfil the (dynamical) Bianchi identity \cite{Blumenhagen:2012nt,Blumenhagen:2012ma,Blumenhagen:2012pc,Geissbuhler:2013uka}
\begin{equation}
\partial_{[A} \mathbf{F}_{BCD]} - \frac{3}{4} {\mathbf{F}^E}_{[AB} \mathbf{F}_{CD]E} = 0, \label{eq:BianchiIdGeneralisedFluxes}
\end{equation}
or in the decomposition into the $d$-dimensional fluxes
\begin{align}
0&= \partial_{[a} \mathbf{H}_{bcd]} - \frac{3}{2} \mathbf{H}_{k[ab} {\mathbf{f}^k}_{cd]} = (\mathrm{d}\mathbf{H})_{abcd} \nonumber \\
0&= \partial^a \mathbf{H}_{bcd} + \partial_{[b} {\mathbf{f}^a}_{cd]} - {\mathbf{f}^a}_{k[a} {\mathbf{f}^k}_{bc]} - \mathbf{H}_{k[bc} {\mathbf{Q}_{d]}}^{ak} \nonumber \\
0&= \partial^{[a} {\mathbf{f}^{b]}}_{cd} + \partial_{[c} {\mathbf{Q}_{d]}}^{ab} - {\mathbf{f}^k}_{ab} {\mathbf{Q}_k}^{cd} + {\mathbf{f}^{[c}}_{m[a} {\mathbf{Q}_{d]}}^{b]k} - {\mathbf{H}}_{abk} \mathbf{R}^{kcd} \label{eq:BianchiIdGeneralisedFluxesDecomp} \\
0&= \partial_a \mathbf{R}^{bcd} + \partial^{[b} {\mathbf{Q}_a}^{cd]} - {\mathbf{Q}_a}^{k[a} {\mathbf{Q}_k}^{bc]} - \mathbf{R}^{k[bc} {\mathbf{f}^{d]}}_{ak} \nonumber\\
0&= \partial^{[a} \mathbf{R}^{bcd]} + \frac{3}{2} \mathbf{R}^{k[ab} {\mathbf{Q}_k}^{cd]}. \nonumber
\end{align}
If the fluxes violate this condition they cannot by written in terms of a generalised vielbein via \eqref{eq:GeneralisedFlux}. In the following we call the corresponding backgrounds \textit{magnetically charged}.

\paragraph{The locally geometric $T$-duality chain and the non-geometric fluxes}

The starting point in the $T$-duality chain is the flat 3-torus with $h$ units of $\mathbf{H}$-flux,  i.e. 
\begin{equation}
\mathbf{H} = h \mathrm{d} x^1 \wedge \mathrm{d} x^2 \wedge \mathrm{d} x^3 .
\end{equation}
A choice of $B$-field for this $\mathbf{H}$-flux is $B =  hx^3 \ \mathrm{d} x^1 \wedge \mathrm{d}x^2$, such that the two commuting isometries of the background are manifest. After a $T$-duality along the isometry $x^1$ the Buscher rules \cite{Buscher:1987sk} produce a pure metric background. This background turns out to be parallelisable, e.g. there is a globally defined frame field ${e_a}^i$. The only non-vanishing component of the generalised flux \eqref{eq:GeneralisedFlux} is
\begin{equation}
{\mathbf{f}^1}_{23} = h \quad \text{with} \quad {\mathbf{f}^c}_{ab} = {e_j}^c {e_{[a}}^i \partial_i {e_{b]}}^j.
\end{equation}
The interpretation of the locally geometric pure $\mathbf{f}$-flux is that it is the totally skewsymmetric combination of the spin connection of a $d$-dimensional vielbein.

Performing a second $T$-duality along $x_2$ we arrive at the background
\begin{align}
G &= \frac{1}{1 + h (x^3)^2} \left(\big(\mathrm{d} x^1 \big)^2 + \big(\mathrm{d} x^2 \big)^2 \right) + \big(\mathrm{d} x^3 \big)^2, \label{eq:HFluxQFlux}\\
 H &= - \frac{h}{\left(1 + h (x^3)^2\right)^2} \left(1 - h(x^3)^2 \right) \mathrm{d} x^1  \wedge\mathrm{d} x^2 \wedge \mathrm{d} x^3 \nonumber
\end{align}
with identifications $x^i \sim x^i + 1$. At $x^3 + 1 \sim x^3$ it is not possible to patch geometrically. Instead we can describe this background by the generalised vielbein
\begin{align}
E_{(Q)} &= \left(\begin{array}{cc} \mathbb{1} & 0 \\ \beta & \mathbb{1} \end{array} \right) , \ {\footnotesize \beta = \left(\begin{array}{ccc} 0 & h x^3 & \\ -h x^3 & 0 & \\ & & 0 \end{array} \right)} \quad \Rightarrow \quad {\mathbf{Q}_3}^{12} = h.
\end{align}
So a constant $\beta$-shift by $h \ \mathrm{d} x^1 \wedge \mathrm{d} x^2$ can be used to patch at $x^3 + 1 \sim x^3$. In other words,  a 3-torus solely together with a constant $\mathbf{Q}$-flux is characterised by a non-trivial monodromy of $\beta$. The background has the interpretation of a non-commutative spacetime with $\PB{x^1}{x^2} \sim hw^3$, where $w^3$ is the winding around the $x^3$-cycle.

Let us summarise this in the scheme \cite{Shelton:2005cf}
\begin{equation}
\mathbf{H}_{123} \overset{T_1}{\longleftrightarrow} {\mathbf{f}_1}^{23} \overset{T_2}{\longleftrightarrow} {\mathbf{Q}_1}^{23} \overset{T_3}{\longleftrightarrow} \mathbf{R}^{123}. \label{eq:TdualityChainFluxes}
\end{equation}
The first two steps can be realised via standard abelian $T$-duality, whereas the last step cannot because background (either described by a generalised metric or generalised vielbeine) does not possess a corresponding isometry for $x^3$. More details on the non-geometric interpretation of these backgrounds can be found in \cite{Blumenhagen:2010hj,Lust:2010iy,Blumenhagen:2011ph,Condeescu:2012sp,Andriot:2012an,Chatzistavrakidis:2012qj,Mylonas:2012pg,Bakas:2013jwa, Plauschinn:2018wbo}.

\paragraph{Local non-geometry and the locally non-geometric $T$-duality chain}

In order to allow for such $T$-dualities along non-isometric direction, we need to allow for the dependence on dual coordinates $\tilde{x}_i$. Derivatives with respect to them are included into $\partial_I = (\partial_i ,\partial^i)$. The dependence of functions on the $2d$ coordinates $X^I = (x^i,\tilde{x}_i)$ is normally restricted by constraints
\begin{align}
\text{strong contraint:} \quad 0&= \partial_I f (X) \cdot \partial^I g(X), \nonumber \\
\text{weak constraint:} \quad 0&= \partial_I \partial^I f(X)  \label{eq:DFTConstraints}
\end{align}
for all functions $f,g$. The strong constraint is typically considered a consistency condition of the gauge algebra of double field theory - as such a consistency or simplifying condition it will also appear in section \ref{chap:DFT} -- whereas the weak constraint corresponds to the level matching condition acting as an operator. Now we can understand $T$-duality simply as the exchange $x^i \leftrightarrow \tilde{x}_i$.

A background is called \textit{locally} non-geometric, if the generalised metric resp. the vielbein depends on the dual coordinates as well. So for example the only choice of a generalised vielbein reproducing a pure $\mathbf{R}$-flux with $\mathbf{R}^{123} = h$ is
\begin{equation}
E_{(R)} = \left(\begin{array}{cc} \mathbb{1} & 0 \\ \beta & \mathbb{1} \end{array} \right), \ {\footnotesize \beta = \left(\begin{array}{ccc} 0 & h \tilde{x}^3 & \\ -h \tilde{x}^3 & 0 & \\ & & 0 \end{array} \right)}.
\end{equation}
For such a background we cannot write down a $\sigma$-model Lagrangian in the usual fashion, as metric and $B$-field do not depend on the coordinates alone.

But we could also choose locally non-geometric generalised vielbeins for a pure $\mathbf{f}$- or $\mathbf{Q}$-flux background, e.g.
\begin{align}
\tilde{E}_{(f)} &= \left(\begin{array}{cc} \mathbb{1} & B \\ 0 & \mathbb{1} \end{array} \right), \ {\footnotesize B = \left(\begin{array}{ccc} 0 & h \tilde{x}^3 & \\ -h \tilde{x}^3 & 0 & \\ & & 0 \end{array} \right) }, \quad \tilde{E}_{(Q)} &= \left(\begin{array}{cc} e & 0 \\ 0 & (e^{T})^{-1} \end{array} \right), \  { \footnotesize e = \left(\begin{array}{ccc} 1 & 0 & 0 \\ -h \tilde{x}^3 & 1 & 0 \\ 0 & 0 & 1 \end{array} \right) }.  \label{eq:LocallyNonGeometricVielbeins} 
\end{align}
It seems impossible to write down a locally non-geometric generalised vielbein for a pure $\mathbf{H}$-flux background, or locally geometric one for a pure $\mathbf{R}$-flux background. The above examples show that local non-geometry is a priori not restricted to $\mathbf{R}$-flux backgrounds.

\paragraph{Examples and Lagrangians}

In the following we will include some explicit examples of such generalised flux frames. Besides setting conventions for later discussion, we would like to emphasise here that in our definition as components of $\bF_{ABC}$ the physical interpretation of the fluxes $\mathbf{H}$, $\mathbf{f}$, $\mathbf{Q}$, $\mathbf{R}$ is \textit{frame dependent}. By this we mean the $\mathbf{Q}$-flux might not correspond to a monodromy of $\beta$ or closed string non-commutativity or the $\mathbf{R}$-flux not to local non-geometry, in a generic generalised frame.

\begin{itemize}
\item \textit{Geometric frame}. This is the standard frame of a Lagrangian $\sigma$-model given by a metric and a $B$-field. Only the $\mathbf{H}$-flux and the geometric $\mathbf{f}$ are non-vanishing
\begin{align}
\mathbf{H}_{abc} &= \partial_{[a}B_{bc]} + {f^d}_{[ab} B_{c]d} \label{eq:GeneralisedFluxesB} \\
{\mathbf{f}^c}_{ab} &= {f^c}_{ab} ={e_j}^c {e_{[a}}^i \partial_i {e_{b]}}^j \nonumber \\
{\mathbf{Q}_c}^{ab} &= 0 = \mathbf{R}^{abc} \nonumber
\end{align}
The corresponding vielbein is a composition of a $d$-dimensional tetrad rotation and a $B$-shift: $E = E^{(B)} E^{(e)}$.

\item \textit{Non-geometric frame }(resp. open string variables). $\sigma$-models like
\begin{equation}
S = -\frac{1}{2} \int \mathrm{d}^2 \sigma \left( \frac{1}{\gamma - \Pi(x)} \right)_{ab} {e_i}^a {e_j}^b \partial_+ x^i \partial_- x^j.
\end{equation}
are described by the vielbein $E = E^{(\beta)}_\Pi E^{(e)}$, with $E^{(\beta)}_\Pi$ denoting a $\beta$-shift by a bivector $\Pi$. This results in the generalised fluxes
\begin{align}
\mathbf{H}_{abc} &= 0\label{eq:GeneralisedFluxesBeta} \\
{\mathbf{f}^c}_{ab} &= {f^c}_{ab}  \nonumber \\
{\mathbf{Q}_c}^{ab} &= {Q_c}^{ab}  = \partial_c \Pi^{ab} + {f^{[a}}_{dc} \Pi^{b]d}  \nonumber \\
\mathbf{R}^{abc} &= R^{abc}  =  \Pi^{d[a} \partial_d \Pi^{bc]} + {f^{[a}}_{de} \Pi^{b|d} \Pi^{|c]e}. \nonumber
\end{align}
This is an important class of backgrounds as this parameterisation is relevant for open strings in NSNS-backgrounds. Also non-abelian $T$-duals, Poisson-Lie $\sigma$-models are of this form.

\item The\textit{ $e$-$B$-$\Pi$-frame}. The next logical step is to introduce a frame in which all the fluxes are non-vanishing. The nearly exclusively used choice in the literature is the generalised flux frame for the $\sigma$-model
\begin{equation}
S = -\frac{1}{2} \int \mathrm{d}^2 \sigma \left( \frac{1}{\frac{1}{\gamma - B(x)} - \Pi(x)} \right)_{ab} {e_i}^a {e_j}^b \partial_+ x^i \partial_- x^j.
\end{equation}
The corresponding generalised vielbein is of the type $E = E^{(\beta)}_\Pi E^{(B)}  E^{(e)}$  and the resulting generalised fluxes are
\begin{align}
\mathbf{H}_{abc} &= \partial_{[a}B_{bc]} + {f^d}_{[ab} B_{c]d} \label{eq:GeneralisedFluxesEBPi} \\
{\mathbf{f}^c}_{ab} &= {f^c}_{ab} +  \mathbf{H}_{abd} \Pi^{de} \nonumber \\
{\mathbf{Q}_c}^{ab} &= {Q_c}^{ab} + \mathbf{H}_{cde} \Pi^{ad} \Pi^{be} = \partial_c \Pi^{ab} + {f^{[a}}_{dc} \Pi^{b]d} + \mathbf{H}_{cde} \Pi^{ad} \Pi^{ae} \nonumber \\
\mathbf{R}^{abc} &= R^{abc} + \mathbf{H}_{def} \Pi^{ad} \Pi^{be} \Pi^{cf} =  \Pi^{d[a} \partial_d \Pi^{bc]} + {f^{[a}}_{de} \Pi^{b|d} \Pi^{|c]e} + \mathbf{H}_{cde} \Pi^{ad} \Pi^{ae}. \nonumber
\end{align}
This has been derived several times in the literature \cite{Roytenberg:2001am,Halmagyi:2008dr,Blumenhagen:2012pc}.

\item The \textit{$e$-$\Pi$-$B$-frame}. The previous choice was not the only possible one. For example $E = E^{(B)} E^{(\Pi)} E^{(e)}$ is a valid parameterisation for which generically all the components of $\bF_{ABC}$ might be non-vanishing. Here the generalised fluxes are
\begin{align}
\mathbf{H}_{abc} &= \partial_{[a}B_{bc]} + {f^d}_{[ab} B_{c]d} + [B,B]^{K.S.}_{abc} + B_{[\underline{a}d} B_{\underline{b}e} {Q_{\underline{c}]}}^{de} + B_{ad}B_{be} B_{ce} \mathbf{R}^{def} \label{eq:GeneralisedFluxesEPiB} \\
{\mathbf{f}^c}_{ab} &= {f^c}_{ab} +  {Q_{[a}}^{dc} B_{b]c} + \Pi^{cd} \partial_d B_{ab} + B_{ab} B_{be} \mathbf{R}^{abc} \nonumber \\
{\mathbf{Q}_c}^{ab} &= {Q_c}^{ab} + \mathbf{R}^{abd} B^{ac}  =  \partial_c \Pi^{ab} + {f^{[a}}_{dc} \Pi^{b]d} + \mathbf{R}^{abd} B^{ac} \nonumber \\
\mathbf{R}^{abc} &=   \Pi^{d[a} \partial_d \Pi^{bc]} + {f^{[a}}_{de} \Pi^{b|d} \Pi^{|c]e}. \nonumber
\end{align}
We recognise the (dual) Koszul derivative $\partial_\Pi^c = \Pi^{cd} \partial_d$, which is used to define the Koszul-Schouten bracket $[ \ , \ ]^{K.S}$ of forms analogously to the usual Schouten bracket of multivector fields. This vielbein corresponds to the $\sigma$-model
\begin{equation}
S = -\frac{1}{2} \int \mathrm{d}^2 \sigma \left( \frac{1}{\gamma^{-1} - \Pi(x)}  - B(x) \right)_{ab} {e_i}^a {e_j}^b \partial_+ x^i \partial_- x^j.
\end{equation}

\item The completely general expression for $\bF_{ABC}$ in terms of a generic generalised vielbein can be found in \cite{Plauschinn:2018wbo}, also including a vielbein which might violate the strong constraint.
\end{itemize}
In contrast to the case in the $T$-duality chain, where only one of the fluxes $\mathbf{H}$, $\mathbf{f}$, $\mathbf{Q}$, $\mathbf{R}$ was turned on, the single components have no general interpretation. E.g. here there can be $\mathbf{R}$-flux in a locally geometric background, if other fluxes are turned on as well.

\paragraph{Global non-geometry}
Metric and $B$-field, encoded in the generalised metric, are only defined locally. If the patching involves only $B$-field gauge transformations and \linebreak $d$-dimensional diffeomorphisms we call the background \textit{globally geometric}. On the other hand for a generic non-geometric background we can patch as
\begin{equation}
\mathcal{H}^\prime_{IJ}(G^\prime(x),B^\prime(x)) = {M_I}^K(x)(  \mathcal{H}_{KL}(G(x),B(x)) {M_J}^L(x)
\end{equation}
for an $M_{KL}(x) \in$O$(d,d)$.  In a corresponding generalised flux frame \eqref{eq:GeneralisedFluxFrame} we have that the 'internal' generalised metric $\mathcal{H}_{AB} = {E_A}^I \mathcal{H}_{IJ} {E_B}^J $ is trivial and globally well-defined. The generalised vielbein will in general be defined only patch-wise and patched via ${E^{\prime A}}_I(x) = {M_I}^J(x) {E^A}_J(x)$. The generalised fluxes transform according to
\begin{equation}
\tilde{\bF}_{ABC} = \bF_{ABC} + {E_{\underline{[A}}}^J {E_{\underline{B}}}^K (\partial_{\underline{C}]} {M^I}_J)M_{IK}. \label{eq:FGaugeTrafo}
\end{equation}
'O$(d,d)$ gauge transformations' are those $M(x)$ for which the second term vanishes such that, as expected, the generalised fluxes are globally defined description in a non-geometric backgrounds. Such O$(d,d)$ gauge transformations include for example
\begin{itemize}
\item  in the \textit{geometric frame}: geometric gauge transformations, i.e. $B$-field gauge transformations and $d$-dimensional diffeomorphisms.
\item  in the \textit{geometric frame with $\mathbf{H} = 0$}: certain (coordinate dependent) $\beta$-shifts in non-holonomic coordinates, s.t. both ${\mathbf{Q}_c}^{ab} = 0$ and $\mathbf{R}^{abc} = 0$. Such $\beta$-shifts exist, homogeneous Yang-Baxter deformations of group manifolds are of this kind for example \cite{Lust:2018jsx}. It has been shown that these correspond to a \textit{non-local} field redefinition in the Lagrangian \cite{Borsato:2016pas}.
\item \textit{frame independent}: all constant O$(d,d)$ transformations, including factorised dualities. For example the constant $\mathbf{Q}$-flux background in the $T$-duality chain is of this type, where $M$ is a constant $\beta$-shift.
\end{itemize}
As \eqref{eq:FGaugeTrafo} shows the allowed ${M_I}^J(x)$ depend on the generalised frame ${E_A}^I$ under investigation. Not all of these necessarily have to be interpretable as standard abelian $T$-duality, for example they might also correspond to non-abelian $T$-duality transformations \cite{Bugden:2019vlj}.

Finding such a generalised flux frame for some given generalised metric $\mathcal{H}$ is non-trivial and not unique, as we have a huge gauge freedom\footnote{Condition \eqref{eq:GeneralisedFluxFrame} fixes only the gauge for the flat internal indices, the gauge freedom corresponds to the gauge freedom of the original $\mathcal{H}_{IJ}$.}. There is in general no preferred frame, except if we can find a globally well-defined generalised vielbein (this case is called a generalised parallelisable manifold - see e.g. \cite{Grana:2008yw}). We are only concerned with local properties of the target space in the following, so all statements involving the generalised vielbeins ${E_A}^I$ are to be understood in a single patch.

In section \ref{chap:Hamiltonian} we strive for a Hamiltonian formulation of classical string theory given directly in terms of these globally well-defined generalised fluxes $\bF_{ABC}$. This formulation will only hide the fact that in principle we still need to work in the different coordinates patches in which the ${E_A}^I$ are defined. Steps towards a more rigorous discussion of global issues have been taken in \cite{Belov:2007qj,Hekmati:2012fb} in  the present context of current algebras and loop groups as phase space.

\subsection{String in an $\mathbf{H}$-flux background}
\label{chap:KalbRamond}

The generalisation of the point particle in a electromagnetic field (section \ref{chap:EM}) to strings in a geometric $\mathbf{H}$-flux background was achieved in \cite{Alekseev:2004np}. We review this result here to set a basis for later discussion. Consider the $\sigma$-model of a (classical) string in a geometric background, defined by metric $G$ and Kalb-Ramond field $B$.
\begin{equation}
S = -\frac{1}{2} \int \mathrm{d} x^i \wedge \left( G_{ij}(x) \star  + B_{ij} \right)  \mathrm{d} x^j 
\end{equation}

\paragraph{Free loop space} The configuration space of a closed string moving in a manifold $M$ is the (free) loop space
\begin{equation}
LM = \left\{ x: \ S^1 \rightarrow M , \ \sigma \mapsto x(\sigma) \right\}. \nonumber
\end{equation}
We denote elements of $LM$ by $x$ or $x^i(\sigma)$, working in a coordinate patch of $M$. We take $\sigma$ to have values between $0$ and $1$ and in a slight abuse of nomenclature for $LM$ also sometimes discuss open strings, by discussion of different boundary conditions on the $x(\sigma)$.

The class of smooth functions on $LM$, that we will consider most often, are (multi-local) functionals on $M$
\begin{equation}
 F: \ LM \rightarrow \mathbb{R}, \ F[x] = \oint \mathrm{d}\sigma_1 ... \mathrm{d}\sigma_n f(x(\sigma_1),...,x(\sigma_n)) \nonumber
\end{equation}
induced by \textit{smooth} functions $f: \ M \times ... \times M \rightarrow \mathbb{R}$ -- in particular this includes all the background fields and fluxes. We assume \textit{no explicit} $\sigma$-dependence required by independence under $\sigma$-reparameterisations.

The tangent space $T(LM)$ is spanned by variational derivatives and consists of elements
\begin{equation}
V[x] = \oint \mathrm{d}\sigma \ V^i[x](\sigma) \frac{\delta}{\delta x^i(\sigma)} \in T(LM). \nonumber
\end{equation}
For simplicity of the notation, we will write $V^i(\sigma) \equiv V^i[x](\sigma)$. These $V^i(\sigma)$ are also only implicit functions of $\sigma$, i.e. $\partial V^i(\sigma) \equiv \int \mathrm{d} \sigma^\prime \partial x^j(\sigma^\prime) \frac{\delta}{\delta x^j(\sigma^\prime)}V^i(\sigma)$, where $\delta =\oint \mathrm{d}\sigma \ \delta x^i(\sigma) \ \frac{\delta}{\delta x^i(\sigma)}$ is the de Rham differential on $LM$ and we use the notation $\partial \equiv \partial_\sigma$.

Not all functions on $LM$ are related to multilocal functionals of smooth functions on $M$. E.g., take the winding number
\begin{equation}
w = \oint \mathrm{d} \sigma \ \partial x(\sigma),
\end{equation}
where $x(\sigma) = x + w \sigma + \ oscillators$, is a total derivative under the integral over the closed circle, but the coordinate $x(\sigma)$ itself is not a \textit{smooth} function on the circle. So not all expressions $\oint \mathrm{d} \sigma \ \partial(...)$ are expected to vanish.

\paragraph{Twisted symplectic structure}

Following the same steps as in section \ref{chap:EM}, we express the symplectic structure in terms of the kinematic momenta $\pi_i := p_i + B_{ij}(x) \partial x^j$
\begin{align}
\omega &= \int \mathrm{d} \sigma \ \delta p_i (\sigma) \wedge \delta x^i(\sigma) \label{eq:SymplecticStructureTwisted} \\
&= \int \mathrm{d} \sigma \left( \delta \pi_i \wedge \delta x^j - \frac{1}{2} \mathbf{H}_{ijk}(x) \partial x^k \delta x^i \wedge \delta x^j + \frac{1}{2} \partial \left(B_{ij}(x) \delta x^i \wedge \delta x^j \right) \right). \nonumber
\end{align}
Up to the total derivative term, the symplectic structure is twisted in a $B$-field gauge independent way, by the $\mathbf{H}$-flux, similarly to the electromagnetic case \eqref{eq:SymplecticStructureEMtwisted}. Imposing that the symplectic form \eqref{eq:SymplecticStructureTwisted} is closed, 
\begin{align}
\delta \omega = \frac{1}{6} \oint \mathrm{d} \sigma \partial_{[i} \mathbf{H}_{jkl]}(x) \ \partial x^i\ \delta x^j \wedge \delta x^k \wedge \delta x^l  = 0, \label{eq:SymplecticStructureTwistedClosure}
\end{align}
requires that $H$ is a closed 3-form on $M$. If we instead neglect the boundary contribution in the symplectic two-form \eqref{eq:SymplecticStructureTwisted}, we get such a contribution for the closure of the symplectic form
\begin{equation}
\delta \omega_{bdy} = \partial \left(\mathbf{H}_{ijk}(x) \delta x^i \wedge \delta x^j \wedge \delta x^k \right)
\end{equation}
up to a total derivative term. So together with  the Hamiltonian
\begin{equation}
H = \frac{1}{2} \oint \mathrm{d} \sigma \left( G^{ij}(x) \pi_i \pi_j + G_{ij}(x) \partial x^i \partial x^j \right) \label{eq:HFluxHamiltonian}
\end{equation}
this defines a world-sheet theory in backgrounds which are magnetically charged under the NSNS flux, e.g. the NS5-brane - in particular in and near these magnetic sources, but requiring that the phase space there is only almost symplectic.

The total derivative terms in \eqref{eq:SymplecticStructureTwisted} resp. \eqref{eq:SymplecticStructureTwistedClosure}, which naively vanish for closed strings, are for example relevant for
\begin{itemize}
\item \textit{open strings ending on $D$-branes}. A contribution to the symplectic structure from a (potentially pure-gauge) $B$-field on the brane is the well known source for the fact, that we find non-commutative gauge theories on the brane. In the present context of deformations of the symplectic/Poisson structure this has been discussed in \cite{Alekseev:2004np}, in particular closedness of the symplectic structure requires $\mathbf{H} \big\vert_{D-\text{brane}} = 0$ if we neglect the boundary term in the current algebra. For the some of the models motivating this article $D$-branes have been discussed, i.e. Poisson-Lie $\sigma$-models \cite{Klimcik:1995np} or $\lambda$-deformations \cite{Driezen:2018glg}.

\item \textit{winding strings}. As discussed above the winding number $w = \oint \mathrm{d} \sigma \ \partial x(\sigma)$ along a compact direction is such an integral over a total derivative. In section \ref{chap:NonGeometry} we show that such winding contributions need to be considered so that the current algebra still satisfies the Jacobi identity.

\item \textit{globally non-geometric backgrounds}. E.g. consider the $\mathbf{Q}$-flux background obtained from the standard $T$-duality chain of $T^3$ with $q$ units of $\mathbf{H}$-flux, expressed in terms of a metric $G$ and the $\mathbf{H}$-flux \eqref{eq:HFluxQFlux}. We expect a contribution of a monodromy $\mathbf{H}(1) - \mathbf{H}(0)$. But let us note that also the Hamiltonian \eqref{eq:HFluxHamiltonian} is not well-defined at $x^3 + 1 \sim x^3$ in the geometric frame.

Choosing the generalised flux frame instead -- here in particular the one for the pure $\mathbf{Q}$-flux background -- should give a globally well-defined description of the background and be used to twist the symplectic structure. Hence this is the route we want to take in the following, in particular in section \ref{chap:Hamiltonian}, and the main results of this article.
\end{itemize} 
It turns out that these twists by the generalised fluxes are more conveniently defined in terms of the variables $p_i(\sigma)$ and $\partial x^i(\sigma)$ and their Poisson structure, the current algebra\footnote{The current algebra and its deformations could in principle also be phrased in terms of a symplectic structure. But in case of such a Poisson structure containing so-called \textit{ultralocal} terms, the symplectic structure will be non- resp. bi-local:
\begin{equation}
\omega_{\text{current}} = \int \mathrm{d}\sigma_1 \mathrm{d}\sigma_2 \bar{\Theta}(\sigma_1 - \sigma_2) \delta p_i(\sigma_1) \wedge \delta (\partial x^i)(\sigma_2),
\end{equation}
where $\bar{\Theta}$ is the step function with $\delta_\sigma \bar{\Theta} = \delta(\sigma)$. Trying to invert the $H$-twisted Poisson current algebra \eqref{eq:SymplecticStructureTwisted} to obtain a twisted $\omega_{\text{current}}$ we get:
\begin{equation}
\omega_{IJ}(\sigma_1,\sigma_2) = \left( \begin{array}{cc} A_{ij}(\sigma_1,\sigma_2) & \delta_i^j \bar{\Theta}(\sigma_1 - \sigma_2) \\
- \delta_i^j \bar{\Theta}(\sigma_1 - \sigma_2) & 0 \end{array} \right) \quad \text{with} \quad \int \mathrm{d}\sigma_2 \partial^2_1 A_{ij}(\sigma_1,\sigma_2) = - H_{ijk} \partial x^k (\sigma_1)
\end{equation}
neglecting boundary terms.                       
}:
\begin{equation}
\PB{\partial x^i(\sigma)}{\partial x^j(\sigma^\prime)} = \PB{p_i(\sigma)}{p_j(\sigma^\prime)} =0, \qquad  \PB{\partial x^i(\sigma)}{p_j(\sigma^\prime)} = \delta^i_j \partial_\sigma \delta(\sigma - \sigma^\prime). \label{eq:CurrentAlgNonCovariant}
\end{equation}
Let us make a the connection between what follows in the next sections and the above twisting of the symplectic structure by the $\mathbf{H}$-flux. Going to kinematic momentum $\pi_j$ the Poisson brackets are
\begin{align}
\PB{\partial x^i(\sigma_1)}{\partial x^j(\sigma_2)} &= 0 ,\qquad \PB{\partial x^i(\sigma_1)}{\pi_j(\sigma_2)} = \delta^i_j \partial_1 \delta(\sigma_1 - \sigma_2). \label{eq:CurrentAlgNonCovariantHtwisted} \\
\PB{\pi_i(\sigma_1)}{\pi_j(\sigma_2)} &= - \mathbf{H}_{ijk}(\sigma_1) \partial x^k(\sigma_1) \delta(\sigma_1 - \sigma_2) + \int \mathrm{d}\sigma \partial\left(B_{ij}(\sigma) \delta(\sigma - \sigma_1) \delta(\sigma - \sigma_2) \right) \nonumber
\end{align}
The Jacobi identity imposes, of course equivalently to \eqref{eq:SymplecticStructureTwistedClosure}, $\partial_{[i} \mathbf{H}_{jkl]} = 0$ and in case we neglect the total derivative term in \eqref{eq:CurrentAlgNonCovariantHtwisted} $ \mathbf{H} \big\vert_{\text{D-brane}} = 0$.

\section{Current algebra and algebroid structures}
\label{chap:CourantAlgebroid}
This sections aims to clarify the relation between the standard current algebra, derived from the canonical Poisson structure, and the Courant algebroid structure first discussed in \cite{Alekseev:2004np}. Different versions of O$(d,d)$-invariant and -covariant current algebras exist in the literature, all of these differ by total derivative terms $\int \mathrm{d}\sigma \ \partial(...)$. 

\subsection{Definitions}
Let us first define our notation and collect some well-known facts about the algebroid structures relevant to us \cite{Mackenzie:1987lg,Courant:1990dm,Siegel:1993th,Siegel:1993bj,Liu:1995lsa,Roytenberg:2001am,Gualtieri:2003dx,Hitchin:2004ut,Ellwood:2006ya,Zabzine:2006uz,Berman:2010is,Plauschinn:2018wbo}. 

\paragraph{Lie algebroid} A vector bundle $E \rightarrow M$ over a manifold $M$ with a \textit{Lie bracket} $[ \ , \ ]_L$, i.e. skew-symmetric and satisfying the Jacobi identity, on the space of sections $\Gamma(E)$ and an chor, a linear map $\rho: \ E \rightarrow TM$, is called a Lie algebroid (over $M$), iff $[ \ , \ ]_L$ together with the anchor $\rho$ satisfies the Leibniz rule
\begin{equation}
[\phi_1 , f \phi_2]_L = \left(\rho(\phi_1)f \right) \phi_2 + f[\phi_1,\phi_2]_L, \qquad \text{for} \ \phi_1,\phi_2 \in \Gamma(E), \ f \in C^\infty(M). \nonumber
\end{equation}
$[ \ , \ ]$ is the Lie bracket on $TM$ and the fact that $\rho$ is a homomorphism of Lie brackets
\begin{equation}
\rho([\phi_1,\phi_2]_L) = \left[ \rho(\phi_1),\rho(\phi_2) \right], \nonumber
\end{equation}
follows from the Leibniz rule.

\paragraph{Courant algebroid} A \textit{Courant algebroid} over a manifold $M$ is a vector bundle $E \rightarrow M$, together with a bracket $[ \ , \ ]_D$ on $\Gamma(E)$, a fibre-wise non-degenerate symmetric bilinear form $\langle \ , \ \rangle_E$ and an anchor $\rho: \ E \rightarrow TM$, satisfying the following axioms:
\begin{align*}
[\phi_1,[\phi_2,\phi_3]_D]_D &= [[\phi_1,\phi_2]_D,\phi_3]_D + [\phi_2,[\phi_1,\phi_3]_D]_D \\
[\phi_1 , f \phi_2]_D &= \left(\rho(\phi_1)f \right) V_2 + f[\phi_1,\phi_2]_D \\
[\phi , \phi]_D &= \frac{1}{2} \mathcal{D} \langle \phi_1 , \phi_2 \rangle \\
\rho(\phi_1) \langle \phi_2 , \phi_3 \rangle &= \langle [\phi_1 , \phi_2]_D , \phi_3 \rangle + \langle \phi_2 , [\phi_1 , \phi_3]_D \rangle
\end{align*}
for $\phi_i \in \Gamma(E)$, $f\in C^\infty(M)$ and the derivation $\mathcal{D}: \ C^\infty(M) \rightarrow E$:
\begin{align*}
\langle \mathcal{D}f,\phi \rangle = \rho(\phi)f.
\end{align*}
We call $[ \ , \ ]_D$ \textit{Dorfman bracket} in the following, it is also called generalised Lie derivative in the literature. From the first to axioms follows that $\rho$ is a homomorphism of brackets. The third axiom implies that $[ \ , \ ]_D$ is not skew-symmetric, the first axiom describes a certain Jacobi identity for this non skew-symmetric bracket.

\paragraph{Skew-symmetric realisation} A Courant algebroid as defined above possesses an equivalent representation via a skew-symmetric bracket
\begin{equation}
[\phi_1 , \phi_2]_C = \frac{1}{2} \left([\phi_1 , \phi_2]_D - [\phi_2 , \phi_1]_D \right) = [\phi_1 , \phi_2]_D - \frac{1}{2} \mathcal{D}\langle \phi_1 , \phi_2 \rangle, \nonumber
\end{equation}
which we call \textit{Courant bracket}. It satisfies modified axioms -- in particular, the Jacobi identity only holds up to a total derivation by $\mathcal{D}$
\begin{equation}
[\phi_1 , [\phi_2,\phi_3]_C ]_C + \ \text{c.p.} = \mathcal{D} \left( \frac{1}{3} \langle [\phi_1,\phi_2]_C,\phi_3 \rangle + \ \text{c.p.} \right).
\end{equation}

\paragraph{The standard Courant algebroid on $TM \oplus T^\star M$} The Courant bracket for sections $\phi = v + \xi \in TM \oplus T^\star M$ is given by
\begin{equation}
[\phi_1,\phi_2]_C = [v_1,v_2] + \mathcal{L}_{v_1} \xi_2 - \mathcal{L}_{v_2} \xi_1 -  \frac{1}{2} \mathrm{d} \left(\xi_2(v_1) - \xi_1(v_2)\right).
\end{equation}
In the following we use the notation $\phi = \phi^I \partial_I$ with $\partial_I = (\partial_i , \mathrm{d}x^i)$ where the action of $\mathrm{d}x^i$ on functions is $\mathrm{d}x^i . f = 0$. Then the coordinate expression for Courant resp. Dorfman bracket is:
\begin{equation}
[\phi_1,\phi_2]_C^I = \phi_{[1}^J \partial_J \phi_{2]}^I + \frac{1}{2}\eta_{JK} \phi^J_{[1}\partial^I\phi_{2]}^K, \qquad [\phi_1,\phi_2]_D^I = \phi_{[1}^J \partial_J \phi_{2]}^I + \eta_{JK} \phi^J_{1}\partial^I\phi_{2}^K, \label{eq:CourantBrackets}
\end{equation}
where $\eta$ is the O$(d,d)$ metric, which raises indices $I,J = 1, ... , 2d$. The anchor is simply projection to $TM$: $v + \xi \mapsto v$.

One motivation for the Courant bracket from the point of view of the study of $T$-dualities is, that it possesses an invariance under global O$(d,d)$-transformations (manifest through the index structure in \eqref{eq:CourantBrackets}) and also under the geometric subgroup of local O$(d,d)$-transformations, namely diffeomorphisms and $B$-field gauge transformations via $\partial_i \rightarrow \partial_i + B_{ij}(x) \mathrm{d}x^j$ with $\mathrm{d}B=0$ .

\subsection{Current algebra as Lie and Courant algebroids}
The current algebra derived from the canonical Poisson structure is given by \eqref{eq:CurrentAlgNonCovariant}. We write it in an O$(d,d)$-covariant way, defining $\bE_I(\sigma) = (p_i(\sigma) , \partial x^j(\sigma))$,
\begin{equation}
\PB{\bE_I(\sigma_1)}{\bE_J(\sigma_2)}[\sigma] = \frac{1}{2} \eta_{IJ} (\partial_1 - \partial_2) \left( \delta(\sigma - \sigma_1) \delta(\sigma - \sigma_2) \right) + \frac{1}{2}\omega_{IJ} \partial \big( \delta(\sigma - \sigma_1) \delta(\sigma - \sigma_2) \big) \label{eq:PBCanonicalCurrent}
\end{equation}
without neglecting the second total derivative term and where we employ the notation: $\PB{ \cdot }{ \cdot } = \oint \mathrm{d}\sigma \left( \PB{ \cdot }{ \cdot }[\sigma] \right)$.\footnote{Here and later in the text we make use of the distributional identities
\begin{align}
(\partial_1 + ... + \partial_n) \big(\delta(\sigma - \sigma_1) \cdot ... \cdot \delta(\sigma - \sigma_n) \big) &=  \partial \big( \delta(\sigma - \sigma_1) \cdot ... \cdot \delta(\sigma - \sigma_n) \big) \label{eq:DistributionalIdentities} \\
\frac{1}{2}e(\sigma_1)\cdot e^{-1}(\sigma_2) (\partial_1 - \partial_2) \big(\delta(\sigma - \sigma_1) \delta(\sigma - \sigma_2) \big) &= \frac{1}{2} (\partial_1 - \partial_2) \big(\delta(\sigma - \sigma_1) \delta(\sigma - \sigma_2) \big) \mathbb{1} - \big((\partial e) \cdot e^{-1}\big)(\sigma) \delta(\sigma - \sigma_1)\delta(\sigma - \sigma_2) \nonumber \\
\big( f(\sigma_2) \partial_1 + f(\sigma_1) \partial_2 \big) \delta (\sigma - \sigma_1)\delta(\sigma - \sigma_2)  &= \big(\partial f(\sigma) \big) \delta(\sigma - \sigma_1) \delta(\sigma - \sigma_2) + \partial\big(f(\sigma) \delta(\sigma - \sigma_1) \delta(\sigma - \sigma_2)\big) \nonumber 
\end{align}
for arbitrary (matrix-valued) functions $e$ and $f$, which hold without any additional boundary terms.} The second term containing $\omega = \left( \begin{array}{cc} 0 & - \mathbb{1} \\ \mathbb{1} & 0 \end{array} \right)$ is a total derivative under the $\sigma$-integral and not invariant under $O(d,d)$-transformations. It is the boundary term that was already discussed in section \ref{chap:KalbRamond}.

\subsubsection{Algebroids over $LM$}
To compare it to the definitions of the previous sections, let us compute the algebra of arbitrary multilocal 'charges'. A section $\phi \in \Gamma(E)$ is given by
\begin{equation}
\phi = \phi[x] = \oint \mathrm{d} \sigma \ \phi^I\big(\sigma\big) \bE_I(\sigma)
\end{equation}
The \textit{Poisson bracket} between these sections $\phi$ is
\begin{align}
\PB{\phi_1}{\phi_2} &= \oint \mathrm{d} \sigma_1 \mathrm{d} \sigma_2 \bE_I(\sigma_1) \left( \phi_{[1}^J(\sigma_2) \frac{\delta}{\delta X^J(\sigma_2)} \phi_{2]}^I(\sigma_1) + \frac{1}{2} \phi^J_{[1}(\sigma_2) \frac{\delta}{\delta X_I(\sigma_1)}\phi_{2]J}(\sigma_2) \right. \nonumber \\ 
&{} \quad  \left. + \frac{\delta}{\delta X_I(\sigma_1)} \frac{1}{2} \left( \omega_{KL} \phi^K_{1}(\sigma_2) \phi^L_{2}(\sigma_2) \right)\right)  \label{eq:CurrentLieBracket}
\end{align}
with $\frac{\delta}{\delta X^I(\sigma)} := \left( \frac{\delta}{\delta x^i(\sigma)} , 0 \right)$. Also, we have a natural anchor map $\rho : E \rightarrow T(LM)$ defined via the Poisson bracket
\begin{equation}
\phi \in \Gamma(E) \mapsto \rho(\phi) = \PB{ \ \cdot \ }{\phi} = \int \mathrm{d}\sigma \phi^i(\sigma) \frac{\delta}{\delta x^i(\sigma)} \in \Gamma\big(T(LM)\big).
\end{equation}
The Leibniz rule follows from the properties of the fundamental Poisson brackets. Also the Jacobi identity
\begin{align}
\PB{\phi_1}{\PB{\phi_2}{\phi_3}} + \ \text{c. p.} = 0
\end{align}
holds \textit{identically}, i.e. without any total derivative terms under the $\sigma$-integrals. For this we have to use $\frac{\delta}{\delta X^I(\sigma)} F \frac{\delta}{\delta X_I(\sigma)} G$ for arbitrary functions $F,G$ on $LM$, which is the \textit{strong constraint} of double field theory on $LM$ and follows here from our definition of $\frac{\delta}{\delta X^I(\sigma)}$. Resultantly the full (multilocal) charge algebra is a not only a Lie algebra as expected, but also a \textbf{Lie algebroid} $(E,\PB{\cdot}{\cdot},\rho)$ over the free loop space $LM$. It is something which could be called standard Lie algebroid of the generalised tangent bundle $(T \oplus T^\star)(LM)$, for which the Lie bracket is the semi-direct product of $TM$, with the Lie bracket and $T^\star M$ for an arbitrary manifold $M$,
\begin{equation}
[\phi_1,\phi_2]_L = [v_1,v_2] + \mathcal{L}_{v_1} \xi_2 - \mathcal{L}_{v_2} \xi_1.
\end{equation}
From \eqref{eq:CurrentLieBracket}, which is written in an O$(d,d)$-covariant way, we see that the Lie algebroid bracket is \textit{not} invariant under O$(d,d)$-transformations due to the presence of the last term containing $\omega$.

There is a natural non-degenerate inner product on $E \rightarrow LM$ induced by the $O(d,d)$-metric $\eta$ on $(T\oplus T^\star)M$:
\begin{equation}
\langle \phi_1 , \phi_2 \rangle = \int \mathrm{d}\sigma \ \eta_{IJ} \phi^I_1 (\sigma) \phi^J_2 (\sigma).
\end{equation}
This product is the canonical bilinear form on $(T\oplus T^\star)LM$. Following the definition $\mathcal{D}$, $\langle \mathcal{D}F , \phi \rangle = \rho(\phi) F$, we find the derivation
\begin{align}
\mathcal{D}F[x] &= \int \mathrm{d}\sigma \ \mathbf{E}_I(\sigma) \frac{\delta}{\delta X_I(\sigma)} F[x] = \int \mathrm{d}\sigma \partial x^i(\sigma) \frac{\delta}{\delta x^i(\sigma)} F[x] \label{eq:CurrentCourantDerivation}\\
&=\int \mathrm{d}\sigma_1 ... \mathrm{d}\sigma_n (\partial_1 + .... + \partial_n) f\left(x(\sigma_1),...,x(\sigma_n)\right) . \nonumber
\end{align}
With the help of these objects we can define the \textbf{standard Courant algebroid} on $(T\oplus T^\star)LM$, for which the Courant resp. Dorfman bracket take the form\footnote{For convenience of the reader we also give all the relevant brackets in a local form  in terms of the basis $\bE_I(\sigma)$ of the current algebra:
\begin{itemize}
\item Brackets: 
\begin{align}
\text{Lie}: \quad \PB{\bE_I(\sigma_1)}{\bE_J(\sigma_2)}[\sigma] &=  \frac{1}{2} \eta_{IJ} (\partial_1 - \partial_2) \big( \delta(\sigma - \sigma_1) \delta(\sigma - \sigma_2)\big)  \nonumber \\
&{} \qquad + \frac{1}{2}\omega_{IJ} \partial \big( \delta(\sigma - \sigma_1) \delta(\sigma - \sigma_2) \big) \label{eq:BasisLieBracket} \\
\text{Courant}: \quad \PB{\bE_I(\sigma_1)}{\bE_J(\sigma_2)}[\sigma] &=  \frac{1}{2} \eta_{IJ} (\partial_1 - \partial_2) \big( \delta(\sigma - \sigma_1) \delta(\sigma - \sigma_2)\big) \label{eq:BasisCourantBracket} \\
\text{Dorfman}: \quad \PB{\bE_I(\sigma_1)}{\bE_J(\sigma_2)}[\sigma] &=  \eta_{IJ} \partial_1 \delta(\sigma - \sigma_1) \delta(\sigma - \sigma_2) \label{eq:BasisDorfmanBracket}
\end{align}

\item Non-degenerate O$(d,d)$-invariant inner product:
\begin{align}
\left\langle \bE_I(\sigma_1) , \bE_J(\sigma_2) \right\rangle[\sigma] = \eta_{IJ} \delta(\sigma - \sigma_1) \delta(\sigma - \sigma_2)
\end{align}
\end{itemize}}:
\begin{align}
\PB{\phi_1}{\phi_2}_C &= \oint \mathrm{d} \sigma_1 \mathrm{d} \sigma_2 \bE_I(\sigma_1) \left( \phi_{[1}^J(\sigma_2) \frac{\delta}{\delta X^J(\sigma_2)} \phi_{2]}^I(\sigma_1) + \frac{1}{2} \phi^J_{[1}(\sigma_2) \frac{\delta}{\delta X_I(\sigma_1)}\phi_{2]J}(\sigma_2) \right) \label{eq:CurrentCourantBracket} \\ 
\PB{\phi_1}{\phi_2}_D &= \oint \mathrm{d} \sigma_1 \mathrm{d} \sigma_2 \bE_I(\sigma_1) \left( \phi_{[1}^J(\sigma_2) \frac{\delta}{\delta X^J(\sigma_2)} \phi_{2]}^I(\sigma_1) +  \phi^J_1(\sigma_2) \frac{\delta}{\delta X_I(\sigma_1)}\phi_{2}(\sigma_2) \right) \label{eq:CurrentDorfmanBracket} 
\end{align}
The anchor $\rho$ is the projection onto $T(LM)$ and we have as for any manifold that the standard Courant algebroid over $LM$ is an \textit{exact} Courant algebroid, as the sequence
\begin{equation}
T^\star(LM) \overset{\rho^T}{\longrightarrow} E \overset{\rho}{\longrightarrow} T(LM)
\end{equation}
is an exact one.

But in contrast to an arbitrary manifold we see that the derivation $\mathcal{D}$ produces a total derivative terms under the $\sigma$-integral, which vanishes for multilocal functional $F[x]$, induced by well-defined \textit{smooth} functions $f: \ M\times .... \times M \rightarrow \mathbb{R}$. But for open string this will give a boundary contribution and even for closed strings topological quantities like winding can arise as discussed in section \ref{chap:KalbRamond} -- e.g. $\int\mathrm{d}\sigma \ \partial x \neq=0$ for a winding string along a compact direction parameterised by $x$. In particular the last term of \eqref{eq:CurrentLieBracket}, which spoiled the $O(d,d)$-invariance should not be neglected.

For any manifold $M$, we have that of the three properties -- skew-symmetry, Jacobi identity and O$(d,d)$-invariance -- each of the three brackets -- Lie, Courant or Dorfman -- satisfy two identically and the third one up to a total derivation term (under  $\mathcal{D}$). In contrast to previous literature we will keep track of the total derivative terms at times in the following. In section \ref{chap:NonGeometry} it is shown a contribution from this total derivation term is indeed necessary to ensure associativity of the subalgebra of zero modes (meaning center of mass coordinate $x$ and momentum $p$, and winding $w$) even in locally geometric backgrounds. This agrees with the discussion in this section, where we expect a violation of the Jacobi identity of the Courant bracket (by a total derivation term), but by assumption a Lie algebroid structure of the phase space. 

\subsubsection{Algebroids over $M$} 
In this paragraph we want to tackle two questions
\begin{itemize}
\item Can we find bundle maps $e_\star$, such that
\begin{equation}
T^\star M \overset{e_\star^T}{\longrightarrow} T^\star(LM) \overset{\rho^T}{\longrightarrow} E \overset{\rho}{\longrightarrow} T(LM) \overset{e_\star}{\longrightarrow} TM
\end{equation}
is a (non-exact) Courant algebroid over $M$ with anchor $e_\star \circ \rho$?

\item Does such a bundle map $e_\star$ also extend to a homomorphism of Lie resp. Courant algebroids? What happens to the total derivation terms?
\end{itemize}
In general these questions seem to go beyond the scope of this article, both for reasons of mathematical rigor - which seems to be required if we consider bundle maps which keep track of more of the 'non-local' structure of the full current algebra - and also for physical reasons - we work in a fully generic background so far, so no mode expansion of the basis $\bE_I(\sigma)$ is available. A stringy expansion of the full current algebra could be an interesting question for further study. A more rigorous study of the current algebra and loop space structure of the phase space can be found in previous literature \cite{Belov:2007qj,Hekmati:2012fb}.

Nevertheless we can find a simple example. Let us consider the  bundle map
\begin{equation}
e^0_\star: \ v = \int \mathrm{d} \sigma_1 ... \mathrm{d}\sigma_n v^i(x(\sigma_1),...,x(\sigma_n))  \frac{\delta}{\delta x^i(\sigma_1)} \quad \mapsto \quad v^i(x) \equiv v^i(x,...,x) \partial_i,
\end{equation}
which is something like the push-forward of the evaluation map of the loop space, $e^0: \ LM \rightarrow M, \ x(\sigma) \mapsto x \equiv x(\sigma_0)$ for some $\sigma_0$.  This bundle map is simply the projection from the loop space phase space to the phase space associated to a point of the string. It induces an anchor $e^0_\star \circ \rho$, as we can show that $e_0$ is an Lie algebra homomorphism. This can be used to view the current algebra as an algebroid over $M$, not only over $LM$.

It extends easily to a complete (Lie resp. Courant) algebroid homomorphism $E \rightarrow (T\oplus T^\star)M$. Consider the generic bracket on $E$,
\begin{align}
\PB{\phi_1}{\phi_2}_{a,b} &= \int \mathrm{d} \sigma_1 \mathrm{d} \sigma_2 \bE_I(\sigma_1) \left( \phi_{[1}^J(\sigma_2) \frac{\delta}{\delta X^J(\sigma_2)} \phi_{2]}^I(\sigma_1) + \frac{1}{2} \phi^J_{[1}(\sigma_2) \frac{\delta}{\delta X_I(\sigma_1)}\phi_{2]J}(\sigma_2) \right. \nonumber \\ 
&{} \quad  \left. + \frac{\delta}{\delta X_I(\sigma_1)} \frac{1}{2} \left( a \ \omega_{KL} + b \ \eta_{KL} \right) \phi^K_{1}(\sigma_2) \phi^L_{2}(\sigma_2) \right) \label{eq:CurrentGenericBracket}
\end{align}
for some $a,b \in \mathbb{R}$, which incorporates all brackets discussed in the previous section. $e^0_\star$ defines a bracket on $(T \oplus T^\star)M$
\begin{equation}
e_\star^0 \PB{\phi_1}{\phi_2}^I =  \PB{e_\star^0 \phi_1}{ e_\star^0\phi_2}^I =  \phi_{[1}^J \partial_J \phi_{2]}^I + \frac{1}{2} \phi^J_{[1} \partial^I \phi_{2]J} +  \frac{1}{2} \partial^I \left( a \ \omega_{KL} + b \ \eta_{KL} \right) \phi^K_{1} \phi^L_{2} \label{eq:PointPhaseSpaceGenericBracket}
\end{equation}
and is a true Courant algebroid homomorphism, but the brackets \eqref{eq:CurrentGenericBracket} differ only by total derivative term under the integral, so they might be argued to be equivalent for sufficiently nice charges for closed strings - but their projections to points  are truly inequivalent. This issue is quite logical because the map $e_\star^0$ does not really correspond to a point-particle limit the string\footnote{An obvious candidate for this would seems to be a bundle map associated to the zero mode projection
\begin{equation}
\bar{e}: \ LM \rightarrow M, \ x(\sigma) = x_0 + \bar{x}(\sigma) \mapsto \int x_0 \equiv \mathrm{d} \sigma x(\sigma) = x_0.
\end{equation}
But the push-forward bundle homomorphism,
\begin{equation}
\bar{e}_\star: \ \phi: \ \int \mathrm{d}\sigma \phi^i(\sigma) \frac{\delta}{\delta x^i(\sigma)} \mapsto \left( \mathrm{d}\sigma \phi^i(\sigma) \right),
\end{equation}
turns out have several issues. Written as such
\begin{itemize}
\item It is conceptually ill-defined, because we add vectors of tangent spaces at different points. We would need to transport them back to $x_0$ before summing them up.
\item It will not be an algebra homomorphism.
\end{itemize}
}, but to a restriction of the total phase space (the current algebra) to a local phase space associated to one point on the string. Total derivative terms correspond to a kind of flux on the string, which adds up to zero for closed strings without winding.

\section{The Hamiltonian realisation of the generalised flux frame}
\label{chap:Hamiltonian}
In the last section we only had a very generic look on aspects of current algebras, valid for arbitrary backgrounds - we did not introduce any dynamics. This section aims to show how the Hamiltonian world-sheet theory in any generalised flux background can be defined by a Hamiltonian of the form of the one of the free string. All the background information is  encoded in a deformation of  the Poisson structure. This deformation of the current algebra will be accounted for by the generalised (geometric and non-geometric) NSNS fluxes, in perfect analogy to the point particle in an electromagnetic field. This generalises the result of \cite{Alekseev:2004np}, reviewed in section \ref{chap:KalbRamond}. Many aspects of this were discussed already in \cite{Halmagyi:2008dr,Halmagyi:2009te} from a Lagrangian point of view and for a certain parameterisation of generalised vielbeins reviewed in section \ref{chap:reviewGeneralisedFluxes}.

\subsection{Hamilton formalism for string $\sigma$-models}
Let us consider a generic string $\sigma$-model coupled to metric and $B$-field of a $d$-dimensional target space
\begin{equation}
S = - \frac{1}{2} \int \left( G_{ij}(x) \ \mathrm{d}x^i \wedge \star \mathrm{d}x^j + B_{ij}(x) \ \mathrm{d} x^i \wedge \mathrm{d} x^j \right).
\end{equation}
Choosing conformal gauge, we find the Hamiltonian to be
\begin{align}
H = \frac{1}{2} \oint \mathrm{d} \sigma \mathcal{H}_{IJ}\left(\sigma\right) \bE^I(\sigma) \bE^J(\sigma)
\end{align}
where $\mathcal{H}_{IJ}(\sigma)$ is the generalised metric \eqref{eq:GeneralisedMetric}, which depends on $\sigma$ via the coordinate dependence of $G$ and $B$. $\bE_I(\sigma) = (p_i(\sigma),\partial x(\sigma))$, where $p_i(\sigma)$ is the canonical momentum, fulfils the canonical current Poisson brackets \eqref{eq:CurrentLieBracket}.

\paragraph{Generalised fluxes in Hamiltonian formalism}

Assume we have a generalised flux frame describing our background, e.g. a generalised vielbein ${E_A}^I(x)$ with
\begin{equation}
{E_A}^I(x) {E_B}^J(x) \mathcal{H}_{IJ}\left( G(x), B(x) \right) = \gamma^{AB} = \left( \begin{array}{cc} \gamma^{ab} & 0 \\ 0 & \gamma_{ab} \end{array} \right),
\end{equation}
where $\gamma^{ab}$ is some convenient flat metric in the signature of the target space, e.g. $\gamma^{ab} = \delta^{ab}$. We could be tempted to phrase the Hamiltonian world-sheet theory also in terms of a new basis of the current algebra: $\bE_A = {E_A}^I \bE_I$. The Hamiltonian is again of the form of a 'free' Hamiltonian:
\begin{align}
H &= \frac{1}{2} \int \mathrm{d} \sigma \ \gamma^{AB} \mathbf{E}_A(\sigma)  \mathbf{E}_{B}(\sigma)
\end{align}
Thus all the information is expected to be encoded in the current algebra. The  redefinition $\bE_A = {E_A}^I \bE_I$ of \eqref{eq:PBCanonicalCurrent} results in the twisted current algebra
\begin{align}
\PB{\mathbf{E}_A(\sigma_1)}{\mathbf{E}_B(\sigma_2)}[\sigma] &=  \frac{1}{2}\eta_{AB} (\partial_1- \partial_2)  \left( \delta(\sigma_1 - \sigma) \delta(\sigma_2-\sigma) \right) + \frac{1}{2} \partial \left( \omega_{AB}(\sigma) \delta(\sigma_1 - \sigma) \delta(\sigma_2-\sigma) \right) \nonumber  \\
&{} \quad - {\bF^C}_{AB}(\sigma) \mathbf{E}_C(\sigma) \delta(\sigma_1 - \sigma) \delta(\sigma_2 - \sigma), \label{eq:PBGeneralFluxes}
\end{align}
with $\bF_{ABC} = (\partial_{[A} {E_B}^I ) E_{C]I}$, or decomposed into the four components $\mathbf{H}$, $\mathbf{f}$, $\mathbf{Q}$ and $\mathbf{R}$:
\begin{align}
\PB{e_{0,a}(\sigma)}{e_{0,b}(\sigma^\prime)} &= - \left( {\mathbf{f}^c}_{ab} (\sigma) e_{0,c}(\sigma) + \mathbf{H}_{abc} (\sigma) e_1^c(\sigma) \right) \delta(\sigma - \sigma^\prime) \nonumber \\
\PB{e_{0,a}(\sigma)}{e_{1}^{b}(\sigma^\prime)} &= - \left( {\mathbf{f}^b}_{ca}(\sigma) e_1^{c}(\sigma) + {\mathbf{Q}_a}^{bc}(\sigma) e_{0,c}(\sigma) \right) \delta(\sigma - \sigma^\prime) - \delta^b_a \partial_{\sigma^\prime}(\sigma - \sigma^\prime) \label{eq:PBGeneralExpl}\\
\PB{e_1^a(\sigma)}{e_1^{b}(\sigma^\prime)} &= - \left( {\mathbf{Q}_c}^{ab}(\sigma) e_1^{c}(\sigma) + \mathbf{R}^{abc}(\sigma) e_{0,c}(\sigma) \right) \delta(\sigma - \sigma^\prime) \nonumber
\end{align}
with $\bE_A(\sigma) = \big( e_{0,a}(\sigma) , e^a_1(\sigma) \big)$. In contrast to the $\eta$-term the total derivative term containing $\omega_{AB}$ is \textit{not} invariant under this change of basis as
\begin{equation}
\omega_{AB}(\sigma) = {\bE_A}^I(\sigma) {\bE_B}^J (\sigma) \omega_{IJ} \neq
\left( \begin{array}{cc} 0 & - \mathbb{1} \\ \mathbb{1} & 0 \end{array} \right)_{AB}
\end{equation}
in general. $e$-transformations leave the $\omega_{IJ}$-term invariant compared to \eqref{eq:PBCanonicalCurrent}, whereas for example a $B$- resp. a $\beta$-shift leads to
\begin{equation}
\omega^{(B)} = \left( \begin{array}{cc} 2B & - \mathbb{1} \\ \mathbb{1} & 0 \end{array} \right) \quad \text{resp.} \quad \omega^{(\beta)} = \left( \begin{array}{cc} 0 & - \mathbb{1} \\ \mathbb{1} &- 2 \beta \end{array} \right).
\end{equation}

\paragraph{Equations of motion}

The Hamilton equations of motion are
\begin{align}
\mathrm{d} \star \be_c + \frac{1}{2} \left( {\mathbf{Q}_c}^{ab} + \mathbf{H}_{cmn} \gamma^{ma} \gamma^{nb} \right) \be_a \wedge \be_b + \frac{1}{2} {\mathbf{f}^{\lbrace a}}_{kc} \gamma^{b \rbrace k} \be_a \wedge \star \be_b &= 0 \label{eq:GeneralisedFluxEOM}\\
\mathrm{d} \be^c + \frac{1}{2} \left( {\mathbf{f}^c}_{ab} + \mathbf{R}^{cmn} \gamma_{ma} \gamma_{nb} \right) \be^a \wedge \be^b + \frac{1}{2} {\mathbf{Q}_{\lbrace a}}^{kc} \gamma_{b \rbrace k} \be^a \wedge \star \be^b &= 0 \label{eq:GeneralisedFluxBianchi}
\end{align}
with one-forms $\be^c = {e_\alpha}^c \mathrm{d} \sigma^\alpha$. In terms of a Lagrangian formulation these correspond to an equation of motion and a world-sheet Bianchi identity. The Hamiltonian formalism does not distinguish between these two 'types' of equations of motion, showing that it is a convenient framework to study dualities.

The equations of motion of the string in an arbitrary locally geometric background can be encoded very conveniently into the O$(d,d)$-covariant form 
\begin{equation}
d\mathcal{E}^A + \frac{1}{2} {\bF^A}_{BC}\mathcal{E}^B \wedge \mathcal{E}^C = 0, \quad \text{with} \quad \mathcal{E}^A := ( \be^a , \star \be_a ) \quad \text{resp.} \quad \mathcal{E}^A = \gamma^{AB} \star \mathcal{E}_B . \label{eq:GeneralisedFluxEOMCov}
\end{equation}
In this form the equations of motion are nothing else than the pullback $\mathcal{E} = x^\star {E}$ of a structure equation for frame fields ${{E}^A}_I(x) \mathrm{d} X^I $  together with the constraint $\star {E}^A = \gamma^{AB} {E}_B$.

\paragraph{Virasoro constraints}

To complete the description of a string theory in a generalised flux background we give the Virasoro constraints and their properties. There is, of course, nothing new to expect - they are a consequence of world-sheet reparameterisation invariance and hold identically. Similarly to the Hamiltonian, the constraints and their properties take the same form as the ones for the string in flat space. This relies solely on the fact that the $\bF_{ABC}$ are totally skew-symmetric. The conservation of the energy-momentum tensor additionally requires the equation of motion as usual. So we can phrase the whole dynamics of a string solely in terms of  the generalised fluxes without referring to the generalised vielbeins.

With the definition $T_{\alpha \beta} = \frac{2}{\sqrt{-h}} \frac{\delta S}{\delta h^{\alpha \beta}}$ and choosing a generalised flux frame $\mathbf{E}_A$ as before, these constraints take the form (in flat gauge on the world-sheet)
\begin{align}
T_{00}(\sigma) &= T_{11}(\sigma) = +\frac{1}{2} \gamma^{AB} \bE_A(\sigma) \bE_B(\sigma) = 0, \nonumber \\
T_{01}(\sigma) &= T_{10}(\sigma) = +\frac{1}{2} \eta^{AB} \bE_A (\sigma) \bE_B(\sigma) = 0. \label{eq:VirasoroGeneralisedFluxFrame}
\end{align}
Moreover, their respective zero modes $H$ and $P$ correspond to world-sheet derivatives $\partial_\tau = \PB{\cdot}{ H}$ and $\partial_\sigma = \PB{\cdot}{P}$. Even if we consider the current algebra with all boundary contributions \eqref{eq:BasisLieBracket}, we get the standard Virasoro algebra
\begin{align}
\PB{T_{\pm\pm}(\sigma_1)}{T_{\pm\pm}(\sigma_2)}[\sigma] &= \pm 2 \left(T_{\pm\pm}(\sigma_1) + T_{\pm\pm}(\sigma_2) \right) \frac{1}{2}(\partial_1 - \partial_2) (\delta(\sigma - \sigma_1)\delta(\sigma - \sigma_2)), \nonumber\\
\PB{T_{\pm\pm}(\sigma_1)}{T_{\mp\mp}(\sigma_2)}[\sigma] &=0. \label{eq:VirasoroAlgebraGeneralisedFluxFrame}
\end{align}
Conservation of the energy momentum tensor holds on-shell \eqref{eq:GeneralisedFluxEOMCov} and for totally skew-symmetric ${\bF}_{ABC}$
\begin{equation}
\partial_+ T_{--}(\sigma) \pm \partial_- T_{++}(\sigma) = \pm {\bF_{ABC}}(\sigma) \gamma^{CD} \bE^A(\sigma) \bE^B(\sigma) \bE_D(\sigma) = 0. \label{eq:EMTensorConservation}
\end{equation}

Let us note, that in the following we continue discuss the unconstrained current algebra. In this way the results in the next sections can be applied to generic $\sigma$-models, not only string ones. For a discussion of Dirac brackets in the current algebra in context of the generalised metric formulation see \cite{Blair:2013noa,Blair:2014kla}.

\paragraph{Deformation of current algebra structure and generalised fluxes}

The approach taken above shows a generalisation of the previously known statement, demonstrated in sections \ref{chap:EM} and \ref{chap:KalbRamond} for point particles in Maxwell background or strings in $\mathbf{H}$-flux backgrounds, that the coupling to these background fields can be encoded in a deformation of the symplectic structure of the phase space -- in contrast to introducing interaction terms in the Lagrangian or the Hamiltonian.  So locally the world-sheet theory in \textit{any} generalised flux background is characterised by 
\begin{equation}
\PB{\bE_A(\sigma_1)}{\bE_B(\sigma_2)} = \eta_{AB} \partial_1 \delta (\sigma_1 - \sigma_2) - {\bF^C}_{AB}(\sigma) \bE_C(\sigma_1) \delta(\sigma_1-\sigma_2) \label{eq:CurrentAlgebraTwisted}
\end{equation}
in terms of the generalised fluxes $\bF_{ABC}$, neglecting total derivative terms, together with a 'free' Hamiltonian $H =  \frac{1}{2} \oint \mathrm{d} \sigma \gamma^{AB} \bE_A(\sigma) \bE_B(\sigma)$ (and similarly the full set of Virasoro constraints).\footnote{From this point of view, we could imagine to generalise to a current algebra twisted by the Weitzenb\"ock connection ${\mathbf{\Omega}^C}_{AB}$ \eqref{eq:WeitzenboeckCon} 
\begin{equation}
\PB{\bE_A(\sigma_1)}{\bE_B(\sigma_2)} = \eta_{AB} \partial_1 \delta (\sigma_1 - \sigma_2) - {\mathbf{\Omega}^C}_{AB}(\sigma) \bE_C(\sigma_1) \delta(\sigma_1-\sigma_2).
\end{equation}
This however seems to be a substantial change in the theory, as the Virasoro algebra \eqref{eq:VirasoroAlgebraGeneralisedFluxFrame} and the conservation of the energy momentum tensor \eqref{eq:EMTensorConservation} relies on the total skewsymmetry of $\bF_{ABC}$.} 

This formulation focuses on the physical content of a background, namely the globally well-defined fluxes opposed to the potentially not globally well-defined objects in the generalised metric formulation. In the case of the point particle in an electromagnetic background or the string in $\mathbf{H}$-flux background, this formulation also seemed to be gauge invariant under $A$- resp. $B$-field gauge transformation. Indeed, all the objects in the twisted current algebra \eqref{eq:CurrentAlgebraTwisted} transform as a tensor under \textit{O$(d,d)$ gauge transformations} $\bE_{A^\prime} \rightarrow {E_{A^\prime}}^A \bE_A $. With O$(d,d)$ gauge transformation, we mean as defined in section \ref{chap:reviewGeneralisedFluxes} precisely those $\bE_{A^\prime}$, under which $\bF_{ABC}$ transforms as a tensor. So all results are expected to take a gauge covariant form, as is usual in the generalised flux formulation of double field theory \cite{Geissbuhler:2013uka}. The Bianchi identity, which will be discussed in the next paragraph, will serve as an example for that.

If we wanted to define the Hamiltonian theory only by means of \eqref{eq:CurrentAlgebraTwisted} and a 'free' Hamiltonian $H$, we need to specify the Poisson brackets between the $\bE_A$ and functions of the phase space as well:
\begin{equation}
\PB{\mathbf{E}_A(\sigma)}{f(x(\sigma^\prime))} = \partial_A f(x(\sigma)) \delta(\sigma-\sigma^\prime)
\end{equation}
with $\partial_A = {E_A}^I \partial_I$ and $\partial_I = (\partial_i , 0 )$ as before.

\paragraph{Bianchi identities and magnetically charged backgrounds}
In analogy to the examples in section \ref{chap:review}, let us show what kind of consistency condition the Jacobi identity of the deformed Poisson brackets implies
\begin{align}
0 &= \PB{\bE_{[A}(\sigma_1)}{\PB{\bE_B(\sigma_2)}{\bE_{C]}(\sigma_3)}}[\sigma] \ + \ \text{c. p.} \label{eq:BianchiIdentity} \\
&= \left( \left( \partial_{[A} \mathbf{F}_{BC]D}(\sigma) + {\mathbf{F}^E}_{D[A} \mathbf{F}_{BC]E}(\sigma)\right) \bE^D(\sigma)  + \bF_{ABC}(\sigma) \partial \right) \delta(\sigma - \sigma_1) \delta(\sigma - \sigma_2) \delta(\sigma - \sigma_3) \nonumber \\
&= \left( \partial_{[A} \mathbf{F}_{BCD]} - \frac{3}{4} {\mathbf{F}^E}_{[AB} \mathbf{F}_{CD]E}\right) \bE^D(\sigma)  \delta(\sigma - \sigma_1) \delta(\sigma - \sigma_2) \delta(\sigma - \sigma_3) \nonumber .
\end{align}
We recognise the Bianchi identity of generalised fluxes \eqref{eq:BianchiIdGeneralisedFluxes} in the last line which takes the form of a covariant derivative \cite{Geissbuhler:2013uka}
\begin{align}
\partial_{[A} \mathbf{F}_{BCD]} - \frac{3}{4} {\mathbf{F}^E}_{[AB} \mathbf{F}_{CD]E} = \bigtriangledown_{[A} \bF_{BCD]} = 0.
\end{align}
This calculation holds exactly, meaning without neglecting total derivative terms, if we start with the full form of  \eqref{eq:PBGeneralFluxes} including the total derivative term there. Instead, we could simplify \eqref{eq:PBGeneralFluxes} to \eqref{eq:CurrentAlgebraTwisted} neglecting the total derivative term as previously done in the $\mathbf{H}$-flux case, see section \ref{chap:KalbRamond} or \cite{Alekseev:2004np}.\footnote{On reason for doing this is that the equations of motion for an open Dirichlet string for example, considering all the boundary terms coming from \eqref{eq:PBCanonicalCurrent}, take the inconvenient form
\begin{equation}
d\mathcal{E}^A(\sigma) + \frac{1}{2} {\bF^A}_{BC}(\sigma)\mathcal{E}^B(\sigma) \wedge \mathcal{E}^C(\sigma) = \frac{1}{2}(\eta^{AB} + \omega
^{AB})\gamma_{BC} \mathcal{E}^C(\sigma_1) \delta(\sigma - \sigma_1) \big\vert_{\sigma_1 = 0}^{\sigma_1 = 1}. \label{eq:GeneralisedFluxEOMCovOpen}
\end{equation} } As a consequence we expect an additional total derivative term in the calculation \eqref{eq:BianchiIdentity} --  similar to the differences between Lie algebroids and Courant algebroid structures discussed in section \ref{chap:CourantAlgebroid}. This is indeed the case, the Jacobi identity implies
\begin{equation}
\frac{1}{2}\int \mathrm{d} \sigma \ \partial \left( \partial_{[A} \omega_{BC]} \right) = 0.
\end{equation}
So, e.g. in the geometric frame and for an open Dirichlet string we have
\begin{equation}
\frac{1}{2} \partial_{[a} \omega^{(B)}_{bc]} \big\vert_\text{D-brane} = \mathbf{H}_{abc} \big\vert_\text{D-brane} = 0
\end{equation}
and all components vanishing as a sufficient condition for associativity of the phase space. This reproduces the boundary contribution to open strings in an $\mathbf{H}$-flux background in the Jacobi identity section as expected in section \ref{chap:KalbRamond} resp. ref. \cite{Alekseev:2004np}.

In full analogy to the point particle in magnetic monopole backgrounds, we expect violations of this Bianchi identity and thus of the Jacobi identity of our current algebra for magnetically charged backgrounds. Such backgrounds like NS5-branes and its $T$-duals have been studied in \cite{Villadoro:2007tb,Geissbuhler:2013uka,Andriot:2014uda} in the generalised flux formulation.\footnote{In \cite{Geissbuhler:2013uka} also the following Bianchi identities/potential source terms have been discussed:
\begin{align*}
\mathcal{J} &= \partial^A \bF_A - \frac{1}{2} \bF^A \bF_A + \frac{1}{12} \bF^{ABC} \bF_{ABC}, \\
\mathcal{J}_{AB} &= \partial^C \bF_{CAB} + 2\partial_{[A} \bF_{B]} - \bF^C \bF_{CAB}
\end{align*}
with $\bF_A = {\mathbf{\Omega}^B}_{BA} + 2 \partial_A d$, where $d$ is the generalised dilaton. We do not expect an appearance of these terms in the classical world-sheet theory, as they do explicitly contain the dilaton and the Weitzenb\"ock connection. Thus we will not consider them in the following. From the side of gauged supergravity both $\bF_{ABC}$ as well as $\bF_A$ are known to correspond to electric gauging parameters \cite{Aldazabal:2011nj,Geissbuhler:2011mx}.} They would source the Bianchi identity like.
\begin{equation}
\partial_{[A} \mathbf{F}_{BCD]} - \frac{3}{4} {\mathbf{F}^E}_{[AB} \mathbf{F}_{CD]E} = \mathcal{J}_{ABCD}. \label{eq:BianchiIdMagneticSources}
\end{equation}
In principle  this implies that inside the magnetic sources the background cannot be described anymore by a generalised vielbeins that gives the generalised flux $\bF_{ABC}$ \eqref{eq:GeneralisedFlux}. This means that in this case we cannot untwist the current algebra and that it is not possible to find a Lagrangian description of the world-sheet theory.\footnote{Phrased in other words, there are no Darboux coordinates to this problem, as the canonical Poisson bracket cannot be used to represent the then non-associative phase space.} Working in the Hamiltonian formalism we still have to specify a generalised vielbein resp. frame, in which all the objects are phrased, although this vielbein will not account for the whole amount of $\bF_{ABC}$.

\subsection{Classical $T$-dualities}

The discovery and examination of (generalised) $T$-dualities followed the path of constructions on the Lagrangian level. A classical proof of a duality is finding that such a construction corresponds to a canonical transformation.\footnote{Demanding that the equations of motions take the same form is not enough. Otherwise, for example, the principal chiral model and the WZW model would be the same, as both equations of motion, as well as Bianchi identities can be arranged to be 
\begin{equation}
\mathrm{d} j + \frac{1}{2} j \wedge j = 0, \quad \mathrm{d} \star j = 0
\end{equation}
The difference lies in the meaning of $j$, which in the language of the Hamiltonian formalism corresponds to different Poisson brackets of the component $j$, see e.g. \cite{Faddeev:1987ph} for details. So only if the equations of motion are the same and the transformation leaves the (canonical) Poisson structure invariant, thus is a canonical one, we can say that the two models are dual to each other.} 

In this section we want to pinpoint peculiarities on $T$-duality from the point of view of the Hamiltonian formalism in the generalised flux frame. We reverse the logic and construct canonical transformations that can be interpreted as (classical) $T$-dualities between different $\sigma$-model Lagrangians.

\subsubsection{(Generalised) $T$-duality}

In a Hamiltonian formulation there is no notion of duality, only the more general notion of canonical transformations. Let us give some criteria from point of view of the generalised flux frame.

\paragraph{On canonical transformations and dualities}
\begin{itemize}
\item A generalised flux $\mathbf{F}_{ABC}$ and a generalised metric $\mathcal{H}$ do \textit{not} yet define a string $\sigma$-model Lagrangian. We need to specify a corresponding generalised vielbein ${E_A}^I$ or in other words Darboux coordinates of our deformed current algebra.

This choice of generalised vielbein might not be unique. Different generalised vielbeins for a given generalised flux background correspond to dual $\sigma$-models Lagrangians.

\item The framework, that we choose to study dualities, are models with \textit{constant} generalised fluxes $\mathbf{F}_{ABC}$. In slight contrast to earlier in this section we define a generic string model in the generalised flux frame by a Hamiltonian defined by a constant generalised metric $\mathcal{H}(G_0,B_0)$.\footnote{If we do not relax the condition $\mathcal{H}=\mathbb{1}$ on the generalised flux frame, the component connected to the identity of O$(d,d)$ will generically lead out of this condition: $M \mathcal{H} M^T \neq \mathbb{1}$.}

The duality group is realised linearly. I.e. a group element ${M_{A^\prime}}^B$ leads to a dual model defined by 
\begin{align}
\mathcal{H}_{{A^\prime}{B^\prime}}(G^\prime_0,B^\prime_0) = {M_{A^\prime}}^C {M_{B^\prime}}^D \mathcal{H}_{CD}(G_0,B_0), \quad \mathbf{F}^\prime_{{A^\prime}{B^\prime}{C^\prime}} = {M_{A^\prime}}^D {M_{B^\prime}}^E {M_{C^\prime}}^F \mathbf{F}_{DEF} \label{eq:GeneralisedTDualitiesDefinition}
\end{align}
Given that we find generalised vielbeins, ${E_A}^I$ resp. ${E^\prime_{A^\prime}}^I(x)$ to the original generalised fluxes $\mathbf{F}_{ABC}$ resp. the dual ones $\mathbf{F}^\prime_{{A^\prime}{B^\prime}{C^\prime}}$, this defines two $\sigma$-model Lagrangians with equivalent Hamiltonian dynamics.

\item The ${M_{A^\prime}}^B$ are O$(d,d)$-matrices, in order to keep the current algebra \eqref{eq:PBGeneralFluxes} form-invariant\footnote{We ignore the non O$(d,d)$-invariant $\omega$-term in \eqref{eq:PBCanonicalCurrent} for our considerations in this section.}. We take them to be constant such that the dual generalised metric and fluxes stay constant.

\item Canonical transformations are normally characterised by generating functions. Our approach instead motivates directly that
\begin{equation}
{M^\prime_I}^J(x) = {E^\prime_I}^{A^\prime}(x) {M_{A^\prime}}^B {E_B}^J(x)
\end{equation}
corresponds a canonical transformation, i.e. leaves the canonical Poisson brackets of the $\bE_I$ \eqref{eq:PBCanonicalCurrent} invariant. In the next section we will motivate the existence of generating functions which would generate exactly to the linearly realised factorised dualities and construct closely related charges on the phase space that generate the component connected to the identity of O$(d,d)$. 

\item From the Hamiltonian point of view a (constant) basis change of the $\bE_A$ does not seem to make any difference on the first sight. The point is that we keep the role of the $(e_{0,a},e_1^a)$ resp. $(p_i,\partial x^i)$ fixed. So e.g. the $\mathbf{f}$- and $\mathbf{H}$-flux always describe the $e_0$-$e_0$ Poisson bracket and so on. Rotating the generalised fluxes around and finding new generalised vielbeins which may depend on the \textit{same} coordinates $x$ is, what we define to be a duality here.\footnote{It is here where the pure $\mathbf{R}$-flux background fails to exist purely geometrically, as we do not find such generalised vielbein only depending on the $x$}

We could have taken the other perspective of rotating our choices of Darboux coordinates, i.e. what of the $\bE_I$ correspond to $p_i$ or $\partial x^i$. In the language of double field theory these would be different solutions to the strong constraint. Both perspectives are of course equivalent.

\item These duality transformations resp. canonical transformations should not be realisable by \textit{purely local} field redefinitions in the $\sigma$-model Lagrangian, otherwise we would call them symmetries.

\end{itemize}
For the remainder of this section we will discuss standard (abelian) $T$-duality and Poisson-Lie $T$-duality from this point of view. Also we propose a generalisation, which we call \textit{Roytenberg duality}, for the case of frames with generic constant generalised fluxes.

\paragraph{Abelian $T$-duality}
The framework for the study of abelian $T$-duality is a background with commuting isometries. Let us choose coordinates, such that the isometries are manifest and ignore the spectator coordinates that do not correspond to isometries.

Such a model is defined by the Hamiltonian
\begin{equation}
H = \frac{1}{2} \int \mathrm{d} \sigma \ \mathcal{H}^{IJ}(G_0,B_0) \bE_I(\sigma) \bE_J(\sigma)
\end{equation}
with constant $G_0$, $B_0$ and -- neglecting the total derivative term from  \eqref{eq:PBCanonicalCurrent} -- 
\begin{equation}
\PB{\bE_I(\sigma_1)}{\bE_J(\sigma_2)}[\sigma] = \eta_{IJ} \partial(\delta(\sigma-\sigma_1) \delta(\sigma - \sigma_2) ).
\end{equation}
Abelian $T$-duality acts via O$(d,d)$-matrices $M$ as ${M_I}^J \bE_J$, leaving the current algebra invariant, but generating new Hamiltonians. Thus the space of dual models is given by the coset $\frac{\text{O}(d,d)}{\text{O}(d) \times \text{O}(d)}$. This can be seen by going to the model with $\mathcal{H}=\mathbb{1}$, where $\text{O}(d) \times \text{O}(d)$-matrices leave the Hamiltonian as well as the canonical current algebra invariant.

\paragraph{Poisson-Lie $T$-duality} 
The case of \textit{Poisson-Lie $T$-duality} \cite{Klimcik:1995ux,Klimcik:1995dy,Klimcik:1995jn}, and included in there also non-abelian $T$-duality \cite{Rocek:1991ps,delaOssa:1992vci,Alvarez:1994np}, is the one with
\begin{equation}
\mathbf{H} = \mathbf{R} = 0, \quad {\mathbf{f}^c}_{ab} = {f^c}_{ab}, \quad {\mathbf{Q}_c}^{ab} = {\barf_c}^{ab}. \label{eq:FluxPLSM}
\end{equation}
The Bianchi identities of generalised fluxes \eqref{eq:BianchiIdGeneralisedFluxesDecomp} reduce to Jacobi identities of the $f$- and $\barf$-structure constants and a mixed Jacobi identity. The algebraic setting is that the generalised fluxes ${\mathbf{F}^C}_{AB}$ correspond to structure constant of a Lie bialgebra $\mathfrak{d}$. A Lie bialgebra is a $2d$-dimensional Lie algebra, with a non-degenerate symmetric bilinear form $\langle \ , \ \rangle$ on $\mathfrak{d}$ given by the O$(d,d)$-metric $\eta$ and two (maximally) isotropic\footnote{meaning $\langle \mathfrak{g},\mathfrak{g} \rangle = 0$.} subalgebras $\mathfrak{g}$ and $\mathfrak{g}^\star$, of which $f$ resp. $\barf$ are the structure constants. Together with the Hamiltonian corresponding to an arbitrary constant generalised metric $\mathcal{H}(G_0,B_0)$ this model is also known under the name $\mathcal{E}$-model in the literature\footnote{named after the operator ${\mathcal{E}_A}^B = \mathcal{H}_{AC}\eta^{CB}$ fulfilling $\mathcal{E}^2 = \mathbb{1}$}.

It is well-known how the corresponding generalised vielbein looks like: It is of the type $E = E^{(\beta)} E^{(e)}$ as discussed in section \ref{chap:reviewGeneralisedFluxes}. The $d$-dimensional vielbein $e$ is given by the components of the Maurer-Cartan forms to the Lie group $G$ associated to the structure constants ${f^c}_{ab}$, ${e_i}^a = (g^{-1} \partial_i g)^a$ where $g$ are $G$-valued fields. $\beta$ is the homogeneous Poisson bivector $\Pi$ on $G$ defined by the dual structure constants ${\barf_c}^{ab}$, fulfilling
\begin{equation}
\Pi(e) = 0, \qquad \partial_c \Pi^{ab}(g) = {\barf_c}^{ab} + {f^{[a}}_{cd} \Pi^{b]d}.
\end{equation}
This bivector $\Pi$ is uniquely determined by such a Lie bialgebra structure. The corresponding $\sigma$-model has the form
\begin{equation}
S \sim \int \mathrm{d}^2 \sigma \left( \frac{1}{\frac{1}{G_0 + B_0} - \Pi(g)}\right)_{ab} (g^{-1} \partial_+ g)^a (g^{-1} \partial_- g)^b.
\end{equation}
Poisson-Lie $T$-duality acts linearly on the deformed current algebra associated to \eqref{eq:FluxPLSM}. This was discovered already in \cite{Sfetsos:1997pi} and discussed in present form already in \cite{Klimcik:1995dy,Klimcik:2015gba,Demulder:2018lmj}. The total factorised duality simply corresponds to $\mathbf{f} \leftrightarrow \mathbf{Q}$, respectively $\mathfrak{g} \leftrightarrow \mathfrak{g}^\star$. The full duality group, which maintains the structure of the generalised fluxes \eqref{eq:FluxPLSM} of the $\mathcal{E}$-model  is discussed in detail in \cite{Lust:2018jsx}. It is the group of different Manin triple decompositions of the Lie bialgebra $\mathfrak{d}$.

At the Lagrangian level, the duality can be realised by considering a 'doubled' $\sigma$-model with target being the Drinfel'd double $\mathcal{D}$ , and then integrating out d.o.f.s corresponding to different (isotropic) subalgebras $\mathfrak{g}^\star$ of the Lie bialgebra $\mathfrak{d}$ \cite{Tyurin:1995bu,Klimcik:1995dy,Driezen:2016tnz}. Other approaches to Poisson-Lie $T$-duality via double field theory and generalised geometry include \cite{Hull:2009sg,ReidEdwards:2010vp,Hassler:2017yza,Hassler:2019wvn}.

\paragraph{Roytenberg duality} Let us consider the generic case: arbitrary \textit{constant} generalised fluxes and a Hamiltonian corresponding to an arbitrary constant generalised metric $\mathcal{H}(G_0,B_0)$. Let us call this case \textit{Roytenberg} model, as a configuration with a generic generic fluxes with non-vanishing $\mathbf{H}$, $\mathbf{f}$, $\mathbf{Q}$ and $\mathbf{R}$ was first considered in \cite{Roytenberg:2001am}. It is not clear, in contrast to the Poisson-Lie $\sigma$-model, how to find a generalised vielbein for a generic choice of constant generalised fluxes. In section \eqref{chap:reviewGeneralisedFluxes} we introduced two choices of generalised vielbeins which generically turn on all of the four generalised fluxes. We consider choices of generalised vielbein which build upon these two and the one of the Poisson-Lie $\sigma$-model:
\begin{itemize}
\item $E_1 = E^{(B)}_{b} E^{(\beta)}_{\beta_0} E^{(\beta)}_{\Pi} E^{(e)}$
\item $E_2 = E^{(\beta)}_{\beta_0} E^{(B)}_{b_0} E^{(\beta)}_{\Pi} E^{(e)}$
\end{itemize}
as before (and want to have constant generalised fluxes). We take $b$ and $\beta_0$ to be constant, $e$ the vielbein of a Lie group $G$ (corresponding to Lie algebra structure constants ${f^c}_{ab}$) and $\Pi(g)$ to be again a homogeneous Poisson bivector on $G$, associated to dual structure constants ${\barf_c}^{ab}$. This choice of $\beta = \beta_0 + \Pi(g)$ in $E_1$ ensures that the resulting $\mathbf{Q}$- and $\mathbf{R}$-flux are constant as wished. The choice of $\beta = \beta_0 + \Pi(g)$ arose as well, if we go to the complete generalised flux frame of the Poisson-Lie $\sigma$-model, i.e. $\mathcal{H}=\mathbb{1}$, see \cite{Lust:2018jsx}. A generalised version of Poisson-Lie $T$-duality, called \textit{affine} Poisson-Lie $T$-duality, taking into account exactly such constant $\beta_0$'s and mapping between different dual choices of $\beta_0$ and $\Pi(g)$ for $B=0$ was considered in \cite{Klimcik:2018vhl}. In the language of the Poisson-Lie $T$-duality group studied in \cite{Lust:2018jsx} these were 'non-abelian $\beta$-shifts'. Let us give the corresponding fluxes and $\sigma$-model Lagrangians for $E_1$,
\begin{align}
\mathbf{H}_{abc} &= b_{[\underline{a}d} b_{\underline{b}e} {\barf_{\underline{c}]}}^{de} - b_{d[a} {f^d}_{bc]} - b_{[\underline{a}d} b_{\underline{b}e} \beta_0^{f[d} {f^{e]}}_{\underline{c}]f}  + b_{ad} b_{be} b_{cf} \mathbf{R}^{def} \nonumber \\
{\mathbf{f}^c}_{ab} &= {f^c}_{ab} -  b_{d[a} {\barf_{b]}}^{dc} + b_{e[a} \beta_0^{d[e} {f^{c]}}_{b]d}+b_{ad} b_{be} \mathbf{R}^{cde} \nonumber \\
{\mathbf{Q}_c}^{ab} &= {\barf_{c}}^{ab} - \beta_0^{d[a} {f^{b]}}_{cd} + b_{cd} \mathbf{R}^{abd} \label{eq:FluxesRoytenbergDuality1} \\
\mathbf{R}^{abc} &= \beta_0^{[\underline{a}d}\beta_0^{\underline{b}e} {f^{\underline{c]}}}_{de} - \beta_0^{d[a} {\barf_d}^{bc]} \nonumber \\
S_1 \quad &\sim \quad \int \mathrm{d}^2 \sigma \ \left( \frac{1}{\frac{1}{G_0 + B_0} - \beta_0 - \Pi(g)} - b \right)_{ab} (g^{-1} \partial_+ g)^a (g^{-1} \partial_- g)^b, \nonumber
\end{align}
and for $E_2$,
\begin{align}
\mathbf{H}_{abc} &=  b_{[\underline{a}d} b_{\underline{b}e}{\barf_{\underline{c}]}}^{de} - b_{d[a} {f^d}_{bc]}  \nonumber \\
{\mathbf{f}^c}_{ab} &= {f^c}_{ab} - b_{d[a} {\barf_{b]}}^{cd} + \beta_0^{cd} \mathbf{H}_{abd} \nonumber \\
{\mathbf{Q}_c}^{ab} &= {\barf_{c}}^{ab} - \beta_0^{d[a} {{f}^{b]}}_{cd} + \beta_0^{e[a} b_{d[e} {\barf_{c]}}^{b]d} + \beta_0^{ad} \beta_0^{be} \mathbf{H}^{cde} \label{eq:FluxesRoytenbergDuality2} \\
\mathbf{R}^{abc} &=  \beta_0^{[\underline{a}d} \beta_0^{\underline{b}e} {f^{\underline{c}]}}_{de} - \beta_0^{d[a} {\barf_d}^{bc]} - \beta_0^{[\underline{a}d} \beta_0^{\underline{e}b} b_{f[d} {\barf_{e]}}^{df} +\beta_0^{ad} \beta_0^{be} \beta_0^{df} \mathbf{H}_{def} \nonumber \\
S_2 \quad &\sim \quad \int \mathrm{d}^2 \sigma \ \left( \left(\frac{1}{\frac{1}{G_0 + B_0} - \Pi(g)} - b \right)^{-1} - \beta_0 \right)^{-1}_{ab} (g^{-1} \partial_+ g)^a (g^{-1} \partial_- g)^b \nonumber
\end{align}
So by construction the identifications
\begin{align}
b \leftrightarrow \beta_0 \quad \text{and} \quad f \leftrightarrow \barf
\end{align}
correspond to the map between the fluxes
\begin{equation}
\mathbf{H} \leftrightarrow \mathbf{R} \quad \text{and} \quad \mathbf{f} \leftrightarrow \mathbf{Q}.
\end{equation}
This would be what we call the \textit{(factorised) Roytenberg duality} in the terminology of \eqref{eq:GeneralisedTDualitiesDefinition}. At the Lagrangian level, the two $\sigma$-models $S_1$ and $S_2$ are (classically) dual to each other with the identifications
\begin{align*}
G_0^{(1)} + B_0^{(1)} &= \frac{1}{G_0^{(2)} + B_0^{(2)}}, \quad \beta_0^{(1)} = b^{(1)}, \quad \beta_0^{(2)} = b^{(1)}\\
f^{(1)} &= \barf^{(2)} \quad \text{and} \quad f^{(2)} = \barf^{(1)}, 
\end{align*}
where the superscript $(i)$ denotes the quantities in $S_i$ and we raised and lowered the indices appropriately.

Using the two generalised vielbeins $E_1$ and $E_2$ to describe these backgrounds, the Roytenberg duality simply seems to be an extension of the Poisson-Lie $T$-duality group. But these vielbeins are probably not the most general description of constant generalised fluxes, so the above example might give just a vague idea, of what a Roytenberg duality is in general and what kind of $\sigma$-model Lagrangians are connected by it. 

The Roytenberg duality group is the full $\frac{\text{O}(d,d)}{\text{O}(d)\times\text{O}(d)}$ rotating the generalised fluxes and is an interesting object of further study. A Lagrangian derivation of this duality might or might not exist. But still the Hamiltonian theory is well-defined as long as the constant generalised fluxes fulfil the Bianchi identities \eqref{eq:BianchiIdGeneralisedFluxesDecomp}. 

Let us close this section with the following remark. There seems to be no difference between abelian and generalised $T$-dualities from the Hamiltonian point of view. We could have viewed the standard $T$-duality chain of section \ref{chap:reviewGeneralisedFluxes} in same fashion\footnote{The only difference is that it includes one non-isometric spectator coordinate.} -- again the true problem continues to be whether we can find appropriate vielbeins to the new fluxes.

\subsubsection{Realisation in the Poisson algebra}
In this section we want to construct the charges that generate infinitesimal O$(d,d)$-transformation in different generalised flux frames. These will show the need for isometries and are closely related to generating functions of the factorised dualities, not only for abelian $T$-duality but also the generalised version discussed above.

\paragraph{Infinitesimal $\mathfrak{o}(d,d)$-transformations via charges}

Let us define the \textit{non-local} charges\footnote{We use $\int^\sigma$ as a formal expression denoting the antiderivative. More precisely we the following procedure
\begin{equation}
\PB{\mathfrak{Q}_{IJ}}{F(\sigma)}= \frac{1}{4} \lim_{\sigma_0 \rightarrow \sigma} \left( \oint \mathrm{d}\sigma^\prime \int^{\sigma^\prime}_{\sigma_0} \mathrm{d}\sigma^{\prime \prime} \PB{\bE_I (\sigma^{\prime \prime}) \bE_J(\sigma^\prime)}{F(\sigma)} \right)
\end{equation}
where it is only important that $\sigma_0 \neq \sigma$. We will come across similar ambiguities later as well, where we will define doubled coordinates $X^I = (x^i,\tilde{x}_i)$ as fundamental fields in the phase space, with $\bE^I = \partial X^I(\sigma)$.}
\begin{equation}
\mathfrak{Q}_{[IJ]} = \frac{1}{2} \oint \mathrm{d}\sigma \int^\sigma \mathrm{d} \sigma^\prime \  \bE_{I}(\sigma^\prime) \bE_{J}(\sigma)
\end{equation}
which generate $\mathfrak{o}(d,d)$-transformations on the $\bE_K(\sigma)$:
\begin{equation}
\PB{\mathfrak{Q}_{[IJ]}}{\bE_K(\sigma)}= \eta_{K[I} \bE_{J]}(\sigma)
\end{equation}
From this and only with help of the Jacobi identity for the $\bE_I(\sigma)$-current algebra it is easy to show that these charges fulfil the O$(d,d)$ Lorentz algebra
\begin{align}
\PB{\mathfrak{Q}_{[IJ]}}{\mathfrak{Q}_{KL}} = \eta_{IK} \mathfrak{Q}_{JL} \ + \ \text{permutations}.
\end{align}
A general infinitesimal O$(d,d)$-transformation
\begin{equation}
M_{IJ} = \mathbb{1} + m_{IJ}, \qquad m \in \mathfrak{o}(d,d),
\end{equation}
on the phase space is generated by $m^{IJ} \mathfrak{Q}_{[IJ]}$. We have not yet made any assumptions on the background, we worked with the canonical Poisson brackets, resp. in the the generalised metric frame.

The action of these charges on functions of the original world-sheet phase space (functions of $x^i(\sigma)$ and $p_i(\sigma)$) is non-local in general. In particular the action of the $\beta$-transformations acts non-locally on functions on the original manifold
\begin{equation}
\PB{\mathfrak{Q}_{ij}}{f(x(\sigma))} = - \tilde{x}_{[i} \partial_{j]} f(x(\sigma)), \qquad \text{with} \quad \tilde{x}_i(\sigma) = \int^\sigma \mathrm{d}\sigma^\prime p_{i}.
\end{equation}
So far $\tilde{x}(\sigma)$ here is a \textit{non-local} variable on the phase space. With the definitions $\partial_I = (\partial_i , \tilde{\partial}^i)$ and $X^I(\sigma) = \left(x^i(\sigma) , \tilde{x}_i(\sigma) \right)$ we have
for functions in terms of this non-local variable $\tilde{x}$
\begin{equation}
\PB{\mathfrak{Q}_{IJ}}{f \big( X(\sigma) \big) } = - X_{[I}(\sigma) \partial_{J]} f\big(X(\sigma)\big)
\end{equation}
in an O$(d,d)$-covariant way. If we instead considered (multi-)local function(al)s on the current phase space, spanned by the $\bE_I(\sigma)$, everything stays (multi-)local
\begin{align}
\PB{\mathfrak{Q}_{IJ}}{f \big(\bE_K(\sigma) \big) } = - \bE_{[I}(\sigma) \frac{\partial}{\partial \bE^{J]}} f(\sigma).
\end{align}
These charges are (in general) not conserved - they do not commute with the Hamiltonian. Instead they generate infinitesimal O$(d,d)$-transformations of the generalised metric as wished, \textit{if} the generalised metric is constant (again neglecting spectator coordinates). So the charges $\mathfrak{Q}_{IJ}$ generate \textit{abelian} $T$-dualities.

\paragraph{'Non-abelian' $\mathfrak{o}(d,d)$-transformations}

$\mathfrak{Q}_{IJ}$ is a tensor under constant O$(d,d)$-trans-formations, but \textit{not} under local ones (due to the integral). So instead we claim that we have natural charges $\mathfrak{Q}_{AB}$ w.r.t. to some generalised vielbein ${E_A}^I(\sigma)$ by the relation
\begin{equation}
\PB{\mathfrak{Q}_{[AB]}}{{\bE_C}(\sigma)} = \eta_{C[A} \bE_{B]}(\sigma). \label{eq:PoincareOdd}
\end{equation}
Such a $\mathfrak{Q}_{AB}$ exists\footnote{The defining relation \eqref{eq:PoincareOdd} is an ODE in $\sigma$}. An implicit realisation for infinitesimal fluxes would be  
\begin{align*}
\mathfrak{Q}_{[AB]}^{\sigma_0} &= \frac{1}{2} \oint \mathrm{d}\sigma \int^\sigma_{\sigma_0} \mathrm{d} \sigma^\prime \  \bE_{[A}(\sigma^\prime) \bE_{B]}(\sigma) + \bar{\mathfrak{Q}}^{\sigma_0}_{AB} \\
\text{with} \quad \delta \bar{\mathfrak{Q}}_{AB} &= \oint \mathrm{d}\sigma \left(\delta \bE^C(\sigma)\right) {M_C}^D(\sigma) \int_{\sigma_0}^\sigma \mathrm{d}\sigma^\prime {(M^{-1})_D}^E (\sigma^\prime) \  \mathbf{\Omega}_{E[A}(\sigma^\prime) \bE_{B]}(\sigma^\prime),
\end{align*} 
where $M(\sigma) = \exp \left( - \int^\sigma_{\sigma_0} \mathrm{d}\sigma^\prime \mathbf{\Omega}(\sigma^\prime) \right) $ and $\mathbf{\Omega}_{AB}={\bF^C}_{AB} \bE_C$. Finding an integrated form of this expression for a generic background seems highly non-trivial. Nevertheless, assuming that the Poisson brackets of the $\bE_A(\sigma)$ fulfil the Jacobi identity, the Lorentz algebra follows directly from \eqref{eq:PoincareOdd}.

What this means is, that for every choice of generalised vielbein ${\bE_A}(\sigma)$ modulo global O$(d,d)$ transformation, there is a representation of O$(d,d)$ acting on the phase space. 

These are simply the linearly realised (infinitesimal) O$(d,d)$ transformations in the generalised flux frame $\bE_A(\sigma)$ of the previous section. The action of constant but infinitesimal O$(d,d)$-matrix $M^{AB} = \mathbb{1} + m^{AB}$, the corresponding O$(d,d)$ transformation is generated by $m^{AB} \mathfrak{Q}_{AB}$ as in the abelian case and similarly the $\beta$-shifts act non-locally (in momentum) on any function $f(x)$. Again these charges are not conserved but generate O$(d,d)$-transformations on a constant generalised metric defining the Hamiltonian.\footnote{We assumed that the algebra of the $\bE_A(\sigma)$ fulfils the Jacobi identity. There might be problems if the background is magnetically charged or, as we will see later, violates the strong constraint.}

\paragraph{Generating function of factorised dualities} 
Factorised dualities are canonical transformations generated by generating functions of type $F[q,Q]$ \cite{Alvarez:1994wj,Sfetsos:1997pi}. For this type of generating function we have
\begin{equation}
\frac{\delta \mathcal{F}}{\delta q} = p \quad \text{and} \quad \frac{\delta \mathcal{F}}{\delta Q} = -P.
\end{equation}
The generating function for abelian $T$-duality is \cite{Alvarez:1994wj}
\begin{equation}
\mathcal{F}[x,\tilde{x}] = - \frac{1}{2} \oint \mathrm{d} \sigma \ \left( \tilde{x} \partial x - x \partial \tilde{x} \right),
\end{equation}
leading as wished to the identifications
\begin{equation}
p = \partial \tilde{x} \quad \text{and} \quad \tilde{p} = \partial x.
\end{equation}
Using the notation of the previous paragraphs this generating function can be written as 
\begin{equation}
\mathcal{F}_{\mathfrak{Q}_{IJ}}[x,\tilde{x}] = - \eta^{IJ} \mathfrak{Q}_{IJ}
\end{equation}
We include the subscript $\mathfrak{Q}_{IJ}$, as we cannot treat the indices of $\mathfrak{Q}_{IJ}$ as tensor indices.

The generating function for generalised $T$-dualities discussed in the previous subsection are in general very difficult to construct explicitly, see e.g. the construction of the generating function of Poisson-Lie $T$-duality in \cite{Sfetsos:1997pi}. With help of the charges $\mathfrak{Q}_{AB}$, for which we do not know the explicit form but know there algebraic properties on the phase space \eqref{eq:PoincareOdd}, we can easily propose a generating function for any 
\begin{equation}
\mathcal{F}_{\mathfrak{Q}_{AB}}[x,\tilde{x}] = - \eta^{AB} \mathfrak{Q}_{AB},
\end{equation}
which generates factorised dualities in any generalised flux frame $\bE_A$.

The generating function and even the associated canonical transformations seem to exist for any background and for any generalised flux frame, independent of whether the background possesses (generalised) isometries or not. The problem of the canonical transformations in non-isometric backgrounds (such that generalised metric or generalised fluxes are functions of $x$ therein) again is, that the dual fields become functions of $\tilde{x}$ non-local functions of the canonical momenta.

\subsection{Strings in double field theory backgrounds}
\label{chap:DFT}

In the previous section we saw that the action of $T$-duality is not well-defined without assuming some isometry. The obstruction was that the dual backgrounds would become functions of new dual coordinates $\tilde{x}_i(\sigma)$. These, being antiderivatives of the canonical momentum densities $p_i(\sigma)$, are not uniquely defined.

A natural approach to this problem is to define $X_I = (\tilde{x}_i , x^i)$ to be the fundamental fields of the phase space. In particular then the generators of infinitesimal $\mathfrak{o}(d,d)$-transformations $\mathfrak{Q}_{IJ}$ are then simply generators of rotations
\begin{equation}
\mathfrak{Q}_{[IJ]} = \frac{1}{4} \oint \mathrm{d} \sigma X_{[I}(\sigma) \bE_{J]}(\sigma).
\end{equation}
This is the approach to double field theory. 

Remarkably it seems, from the point of view of the Hamiltonian formalism, we would not need to 'double' phase space but instead allow a dependence of the background on the momenta in this very peculiar non-local way, namely via $\tilde{x}_i = \int^\sigma \mathrm{d}\sigma^\prime p_i(\sigma)$.

\paragraph{Poisson brackets on doubled space and the strong constraint}

The question is what the Poisson structure on the doubled space is, and in particular if it is a Poisson structure, i.e. if the proposed brackets fulfils the Jacobi identity. First, we look for a skewsymmetric Poisson bracket $\PB{X_I(\sigma)}{X_J(\sigma^\prime)}$ by integrating the canonical current algebra \eqref{eq:PBCanonicalCurrent}. The solution is
\begin{equation}
\PB{X_I(\sigma)}{X_J(\sigma^\prime)} = - \eta_{IJ} \bar{\Theta}(\sigma - \sigma^\prime) \quad (+ \ c \ \omega_{IJ}) \label{eq:PBDoubledSpace}
\end{equation}
with $\bar{\Theta}(\sigma) = 1/2 \ \text{sign}(\sigma)$, s.t. $\partial_\sigma \bar{\Theta}(\sigma) = \delta(\sigma)$ and an integration constant $c$. As a bracket of functionals,
\begin{equation}
\PB{\Psi_1}{\Psi_2} = \oint \mathrm{d}\sigma_1 \mathrm{d} \sigma_2 \bar{\Theta}(\sigma_1 - \sigma_2) \eta^{IJ} \frac{\delta \Psi_1}{X^I(\sigma_1)} \frac{\delta \Psi_2}{X^I(\sigma_2)},
\end{equation}
it will always vanish if we assume the \textit{strong constraint} (or section condition) of double field theory
\begin{equation}
\eta^{KL} \partial_K f (X) \partial_L g (Y) = 0, \qquad \text{for all } X=(x,\tilde{x})\text{ and }Y=(y,\tilde{y}) \quad \text{and all } f,g. \label{eq:StrongConstraint}
\end{equation}
The bracket \eqref{eq:PBDoubledSpace} without the $\omega$-term has been discussed already in \cite{Blair:2013noa,Blair:2014kla} from point of view of first order $\sigma$-models and the generalised metric formulation. The occurence of the constant $\omega$-term reminds of the zero mode non-commutativity observed in \cite{Freidel:2017wst}.

It is easy to show that the above bracket satisfies the Jacobi identity on the space of functionals (with or without the $\omega$-term). But let us note that the bracket \eqref{eq:PBDoubledSpace} is \textit{not} equivalent to the canonical current algebra, when we are not neglecting $\oint \mathrm{d} \sigma \ \partial(...) \neq 0$ terms. For example trying to derive \eqref{eq:PBCanonicalCurrent} from \eqref{eq:PBDoubledSpace} leads to ambiguities because $\partial_\sigma \partial_{\sigma^\prime} \PB{X_I(\sigma)}{X_J(\sigma^\prime)}$ and $ \partial_{\sigma^\prime} \partial_\sigma \PB{X_I(\sigma)}{X_J(\sigma^\prime)}$ differ exactly by such a topological/total derivative term. 

Accepting this we will use the fundamental brackets
\begin{equation}
\PB{X_I(\sigma)}{X_J(\sigma^\prime)} = - \eta_{IJ} \bar{\Theta}(\sigma - \sigma^\prime), \qquad \qquad \PB{X_I(\sigma)}{\bE_J(\sigma^\prime)} = \eta_{IJ} \delta(\sigma - \sigma^\prime) \label{eq:PBDoubledSpace2}
\end{equation}
supplemented by the canonical current algebra \eqref{eq:PBCanonicalCurrent} for our calculations of the brackets of functionals $F[X,\bE]$, which do not contain $\sigma$-derivatives of $X$ or $\bE$. In section \ref{chap:CourantAlgebroid} it was shown that the Jacobi identity for sections of the 'canonical Lie algebroid' holds exactly when using the canonical current algebra (and also assuming the strong constraint). Allowing for violations of the strong constraint leads to a fundamental violation of the Jacobi identity. As an example let us compute the Jacobi identity between a functional $\Psi[X]$ and two sections of the canonical Lie algebroid $\phi_i[X,\mathbf{E}] = \oint \mathrm{d} \sigma \phi^I_i[X](\sigma) \bE_I(\sigma)$, where $\Psi$ and the $\phi_i^I(\sigma)$ are assumed to be functional of $X(\sigma)$ only, not of its $\sigma$-derivatives. Using \eqref{eq:PBDoubledSpace2} we arrive at
\begin{align}
\PB{\Psi}{\PB{\phi_1}{\phi_2}} + c.p. &= \oint \mathrm{d} \sigma_1 \mathrm{d}\sigma_2 \left[ \frac{1}{2} (\eta_{JK} + \omega_{JK}) \phi_{[1}^J(\sigma_1) \frac{\delta \Psi}{\delta X^I(\sigma_2)} \frac{\delta \phi_{2]}^K(\sigma_1)}{\delta X_I (\sigma_2)} \right. \label{eq:JacobiIdNoSC}\\
&{} \quad + \bE_I(\sigma_1) \bE_J(\sigma_2) \left(\PB{\Psi}{\PB{\phi_1^I(\sigma_1),}{\phi_2^J(\sigma_2)}} + c.p. \right)  \nonumber \\
&{} \quad +  \bE_I(\sigma_1) \left( \scriptstyle{\frac{\delta \phi_{[1}^I(\sigma_1)}{\delta X^J(\sigma_2)} \PB{\Psi}{\phi_{2]}^J(\sigma_2)} - \frac{\delta \Psi}{\delta X^J(\sigma_i)} \PB{\phi_{[1}^I(\sigma_1)}{\phi_{2]}^J (\sigma_2)} - \phi_{[1}^J(\sigma_2) \PB{\Psi}{\frac{\delta \phi_{2]}^I(\sigma_1)}{\delta X^J(\sigma_2)}}}  \right. \nonumber \\
&{} \quad + \left. \left. \scriptstyle{ \frac{1}{2} (\eta_{JK} + \omega_{JK}) \ \phi_{[1}^J (\sigma_2) \PB{\Psi}{\frac{\delta \phi_{2]}^K(\sigma_2)}{\delta X^I(\sigma_1)}} - \frac{1}{2} (\eta_{JK} - \omega_{JK}) \frac{\delta \phi_{[1}^J(\sigma_2)}{\delta X^I(\sigma_1)} \PB{\Psi}{\phi_{2]}^K(\sigma_2)} } \right)  \right]. \nonumber
\end{align}
We recognise the Jacobi identity of the (here unspecified) $X$-$X$ Poisson bracket in the second line of \eqref{eq:JacobiIdNoSC}, which we assume to vanish. Apart from that we see other strong constraint violating term contributing generically to a violation of the Jacobi identity. It might seem that taking different choices of the topological term in the canonical current algebra or a different choice $X$-$\bE$-Poisson bracket than \eqref{eq:PBDoubledSpace2} could make these contributions disappear. But in section \ref{chap:NonGeometry} we will show that this is not the case and fundamental violation (in particular the first term) leads exactly to typical non-vanishing Jacobiators of the zero modes in generalised flux backgrounds. For this we only need the canonical Poisson brackets from which we also derived the current algebra including the topological term.

\paragraph{The generalised flux frame and the Virasoro algebra}
Let us now briefly mention differences, which would occur in the approach taken in section \ref{chap:Hamiltonian}.1, if we allow the generalised vielbeins itself violate the strong constraint (which is in fact not the case in the typical examples of non-geometric backgrounds -- see e.g. the $\mathbf{R}$-flux backgrounds). Going to a generalised flux frame ${E_A}^I(X)$, allowing for a generic dependence on the doubled space, we get the Poisson current algebra
\begin{align}
&{} \quad \PB{\bE_A(\sigma)}{\bE_B(\sigma^\prime)}= \PB{{E_A}^I\big(X(\sigma)\big) \bE_I(\sigma)}{{E_B}^J\big(X(\sigma^\prime)\big) \bE_J(\sigma^\prime)} \label{eq:DFTPB} \\
&= \eta_{AB} \partial_\sigma \delta(\sigma - \sigma^\prime) - {\bF^C}_{AB}(\sigma) \bE_C (\sigma)   \delta(\sigma - \sigma^\prime) - {\mathbf{G}_{AB}}^{CD}(\sigma,\sigma^\prime) \bE_C(\sigma) \bE_D(\sigma^\prime) \bar{\Theta}(\sigma - \sigma^\prime) \nonumber
\end{align}
The last term is bilocal and vanishes if the generalised satisfies the strong constraint. It is given in terms of the Weitzenb\"ock connection \eqref{eq:WeitzenboeckCon}
\begin{align*}
{\mathbf{G}_{AB}}^{CD}(\sigma,\sigma^\prime) =  \eta^{KL}  \left( \partial_K {E_A}^I (\sigma)\right)\left( \partial_L {E_B}^J (\sigma^\prime)\right){E_I}^C(\sigma) {E_J}^D(\sigma^\prime)  = \eta^{KL} {\bOmega_{K,A}}^C(\sigma) {\bOmega_{L,B}}^D(\sigma^\prime).
\end{align*}
If $\bF$ and $\mathbf{G}$ are given in terms of generalised vielbeins, then \eqref{eq:DFTPB} is a Poisson bracket because \eqref{eq:PBDoubledSpace} is. The equation of motion of the string in a DFT background would be also modified by the non-local and strong constraint violating $\mathbf{G}$-terms. 

Also, and more crucially, this term would be responsible for a modification of the Virasoro algebra. Following the derivation of the Virasoro algebra in the generalised flux frame in \ref{chap:Hamiltonian}.1, this can be easily seen as the $\mathbf{G}$-term is not totally antisymmetric. This connection of strong constraint and the algebra of worldsheet diffeomorphisms was already noted from the generalised metric point of view in \cite{Blair:2013noa}. Nevertheless, this only occurs if the generalised vielbein or generalised metric theirselves depend on coordinates and their dual at the same time.

As in section \ref{chap:Hamiltonian}, we calculate the Bianchi identity of the objects $\bF$ and $\mathbf{G}$ by imposing the Jacobi identity on \eqref{eq:DFTPB}
\begin{equation}
\partial_{[A} \bF_{BCD]} - \frac{3}{4} {\bF^E}_{[AB} \bF_{CD]E} = \ \text{strong constraint violating terms}. \label{eq:DFTBianchi}
\end{equation}
Thus one way to account for a violation of the Bianchi identities of generalised fluxes \eqref{eq:GeneralisedFluxBianchi}, e.g. in order to describe magnetically charged backgrounds, is to consider violations of the strong constraint, but only if we trade off (manifest) locality of the equations of motion for it and a modification of the Virasoro algebra for it.

\section{Non-geometry and the deformed current algebra}
\label{chap:NonGeometry}

In this section we want to demonstrate the approach taken in the last section and clarify it by studying examples. We do so by reproducing standard results  for constant a $B$-field in case of the open string, and constant $\mathbf{H}$-, $\mathbf{f}$-, $\mathbf{Q}$- or $\mathbf{R}$-flux for the closed string. The key points are
\begin{itemize}
\item Leaving magnetic or locally non-geometric backgrounds aside, there should be 'Darboux coordinates' $(x^i(\sigma), \ p_i(\sigma))$ fulfilling the canonical Poisson brackets. The question is where the well-known non-geometric nature of the backgrounds is 'hidden', meaning their non-commutative and non-associative behaviour.

In section \ref{chap:Hamiltonian} we saw that beside Darboux coordinates $x^i(\sigma), \ p_i(\sigma)$, the generalised flux frame of a given background gives rise to a second preferred set of coordinates for the current algebra $\bE_A(\sigma)$. We define 'non-geometric coordinates' $y^a$ and 'non-geometric momenta' $\pi_a$ by $\partial y^a = \bE^a(\sigma)$ and $\pi_a(\sigma) = \bE_a(\sigma)$. In the spirit of section \ref{chap:EM} we dub them \textit{'kinematic'}. Their Poisson brackets agree with the known ones usually associated to non-geometric backgrounds. 

With this we can generalise the non-geometric interpretation to more complicated generalised flux backgrounds. Also, we do not need to know the mode expansions of the fields $y^a(\sigma)$ (or impose the equations of motion) to study the non-geometric behaviour of the background.

\item In the spirit of generalised geometry and double field theory, we demonstrate in the language of the current algebra, how $T$-dualities can be reproduced by choosing different solutions to the strong constraint.

\item The significance of the non O$(d,d)$-invariant boundary term in \eqref{eq:PBCanonicalCurrent} or \eqref{eq:PBGeneralFluxes} lies
\begin{itemize}
\item reproducing non-commutativity for the endpoints of open strings.

\item ensuring associativity for closed strings, unless we calculate the brackets of objects violating the strong constraint. In that case, the zero modes of the current algebra (and its integrated form) show that this approach reproduces the known form of non-vanishing Jacobiators in the constant $\mathbf{Q}$- and $\mathbf{R}$-flux backgrounds.
\end{itemize}
\end{itemize}

\subsection{Open string non-commutativity}
In this section we review the classic result of \cite{Chu:1998qz,Seiberg:1999vs} and are interested in the world-sheet dynamics of an open string in a constant $B$-field background. It can be expressed the open string variables resp. the non-geometric frame with flat metric and $\beta$
\begin{equation}
\beta^{ij} = \left(\frac{1}{G+\mathcal{F}}\right)^{ik} \mathcal{F}_{kl} \left(\frac{1}{G-\mathcal{F}}\right)^{lj}, \qquad \mathcal{F} = B - \mathrm{d}A = B-F,
\end{equation}
where $G$ is the flat Minkowski metric and $F$ is the constant field strength of a Maxwell field. The current algebra in the generalised flux basis (the non-geometric frame) in which we have the 'free' Hamiltonian is
\begin{align}
\PB{e_{0,i}(\sigma_1)}{e_{0,j}(\sigma_2)} &= 0 \nonumber \\
\PB{e_{0,i}(\sigma_1)}{e_1^j(\sigma_2)} &= - \delta_i^j \partial_2 \delta(\sigma_1 - \sigma_2)  \label{eq:pureGaugeBetaCurrentAlgebra} \\
\PB{e_{1}^i(\sigma_1)}{e_1^j(\sigma_2)} &= - \beta^{ij} \int \mathrm{d}\sigma \ \partial \left( \delta(\sigma - \sigma_1) \delta(\sigma - \sigma_2) \right). \nonumber
\end{align}
Now we associate new 'non-geometric coordinates' to this new basis: meaning $e_1^a = \partial y_a(\sigma)$. Simply integrating both sides of the last line of \eqref{eq:pureGaugeBetaCurrentAlgebra} gives the result:
\begin{equation}
\PB{y^i(\sigma_1)}{y^j(\sigma_2)} = \left\{ \begin{array}{cc} -\beta^{ij}, & \sigma_1=\sigma_2 = 1 \\ +\beta^{ij}, & \sigma_1 = \sigma_2 = 0 \\ 0 & \text{else.} \end{array} \right.
\end{equation}
This is exactly the result of \cite{Chu:1998qz}, derived without any reference to a mode expansion. Let us note that the total derivative $\omega$-term in the last line of \eqref{eq:pureGaugeBetaCurrentAlgebra} was crucial for this result.

\subsection{Closed string non-commutativity and non-associativity}

Next let us demonstrate the logic explicitly for the well-known standard example of the $T$-duality chain of the 3-torus with constant $\mathbf{H}$-flux. 

First let us consider the $\mathbf{Q}$-flux background ${\mathbf{Q}_3}^{12} = h$, all other components being zero, which is described by the generalised vielbein 
\begin{align}
E_{(Q)} &= \left(\begin{array}{cc} \mathbb{1} & 0 \\ \beta & \mathbb{1} \end{array} \right) , \quad \beta^{12} = h x^3 .
\end{align}
The corresponding current algebra including boundary terms is
\begin{align}
\PB{e_{0,a}(\sigma_1)}{e_{0,b}(\sigma_2)} &= 0 \nonumber \\
\PB{e_{0,a}(\sigma_1)}{e_1^b(\sigma_2)} &= - \delta_a^b \partial_2 \delta(\sigma_1 - \sigma_2) - {\mathbf{Q}_a}^{bc} e_{0,c}(\sigma_1) \delta(\sigma_1 - \sigma_2) \label{eq:QFluxCurrentAlgebra} \\
\PB{e_1^a(\sigma_1)}{e_1^b(\sigma_2)} &= - {\mathbf{Q}_c}^{ab} e_1^c(\sigma_1) \delta(\sigma_1 - \sigma_2) - \int \mathrm{d}\sigma \ \partial \left( \beta^{ab}(\sigma) \delta(\sigma - \sigma_1) \delta(\sigma - \sigma_2) \right). \nonumber
\end{align}
Let us consider the zero modes of the 'kinematic coordinates' associated to this generalised flux frame
\begin{alignat}{2}
p_a &= \oint \mathrm{d} \sigma \ p_a(\sigma) = \oint \mathrm{d} \sigma \ e_{0,a}(\sigma) \qquad \tilde{y}_a &&= \oint \mathrm{d}\sigma^\prime \int^{\sigma^\prime} \mathrm{d} \sigma \ p_a(\sigma) \nonumber \\
w^a &= \oint \mathrm{d}\sigma \ \partial y^a(\sigma) = \oint \mathrm{d}\sigma \ e_1^a(\sigma) \qquad y^a &&= \oint \mathrm{d}\sigma^\prime \int^{\sigma^\prime} \mathrm{d}\sigma \ \partial y^a(\sigma). \nonumber
\end{alignat}
These modes have a priori nothing to do with the original target space interpretation. This seems particular confusing in case of the winding number. But cases like this exist in the literature, there it is sometimes called 'twisted boundary conditions', see e.g. in the context of $\beta$-deformations of $AdS_5 \times S^5$ \cite{Frolov:2005dj}. In the present case we have
\begin{equation}
\partial y^a (\sigma) \equiv e_1^a (\sigma) = \delta^a_i \partial x^i(\sigma) + \beta^{ab}\delta_b^j p_j(\sigma) \label{eq:FieldRedefinition}
\end{equation}
and
\begin{align}
w^3 = w^3_x \qquad \text{and} \qquad w^{1/2} = w^{1/2}_x \pm h \oint \mathrm{d}\sigma x^3 p_{2/1}.
\end{align}
The winding along the $y^3$ direction coincides with the actual one along the $x^3$ direction as also $y^3$ coincides with $x^3$ up to a constant. Now we can integrate the current algebra \eqref{eq:QFluxCurrentAlgebra}.  We use a schematic mode expansion of the kinematic coordinates
\begin{equation}
y^a(\sigma) = y^a + \left( w^a - \frac{1}{2} y^a \right) \sigma + y_{osc}^a(\sigma) \label{eq:ModeExpansion}
\end{equation}
with $y_{osc}^a(\sigma) = y_{osc}^a(\sigma + 1)$ denoting oscillator terms, of which we will not keep track explicitly as we are interested in the zero modes. Alternatively we could approach the this calculation by interserting the most general modes expansions  or $x(\sigma)$ and $p(\sigma)$, that are compatible with the boundary condition, and calculating the contributions of all the modes directly by using the field redefinition \eqref{eq:FieldRedefinition}. This calculation also shows that all the oscillators of the $y$-expansion would still commute with the zero modes, such that they do not give a contribution to the Jacobi identity of the zero modes.

Integrating \eqref{eq:QFluxCurrentAlgebra} the non-vanishing Poisson brackets of the zero modes are
\begin{alignat}{3}
\PB{y^1}{y^2} &\sim - h w^3 + \ osc., && \PB{w^1}{w^2} &&= -h w^3 \nonumber \\
\PB{\tilde{y}_3}{y^1} &\sim - h p_2 + \ osc., &&  \PB{\tilde{y}_3}{y^2} &&\sim h p_1 \ + \ osc. \nonumber \\
\PB{p_3}{w^1} &= -hp_2, && \PB{p_3}{w^2} &&= h p_1 \label{eq:zeromodePBQflux} \\
\PB{y^a}{p_b} &= \delta^a_b + \ osc. , && \PB{\tilde{y}_a}{w^b} &&= \delta_a^b + \ osc. \nonumber \\
\PB{y^1}{w^2} &= \PB{y^2}{w^1} = - h\left(\underline{  y^3 + \frac{1}{2} w^3} + \ osc. \right),  &&  \nonumber
\end{alignat}
reproducing the known non-commutative interpretation of the pure $\mathbf{Q}$-flux background. The underlined terms only stem from the boundary term and $\sim$ denotes some neglected constant factors, including integration constants. Also let us emphasise again that our assumptions do not imply anything about a mode expansion apart from \eqref{eq:ModeExpansion} resp. the boundary conditions. So we can discuss the non-geometric structure without solving the theory first.  

\paragraph{Non-associativity}
There are non-trivial Jacobi identities of the zero mode Poisson brackets:
\begin{align}
\PB{\tilde{y}_3}{\PB{w^1}{w^2}} + \ c.p. \sim \PB{\tilde{y}_3}{\PB{y^1}{y^2}} + \ c.p. \sim h &= {\mathbf{Q}_3}^{12} \\
\PB{w^1}{\PB{y^2}{p_3}} + \PB{p_3}{\PB{w^1}{y^2}} + \PB{y^2}{\PB{p_3}{w^1}} &\sim 0, \nonumber \label{eq:QFluxNA}
\end{align}
neglecting oscillator terms. The zero mode part of the second line vanishes due to the boundary term contribution (the underlined term in \eqref{eq:zeromodePBQflux}). The first line is a non-associativity coming from a potential violation of the strong constraint. In fact it is exactly the expected contribution from the discussion in section \ref{chap:Hamiltonian}.3. Specifying the general expression \eqref{eq:JacobiIdNoSC} of the violation of the Jacobi identity due to strong constraint violations
to $X_I(\sigma)$ and $\bE_A(\sigma)$ gives
\begin{equation}
\PB{X_I(\sigma_1)}{\PB{\bE_A(\sigma_2)}{\bE_B(\sigma_3)}} + c.p. = \frac{1}{2} \left( \eta_{MN} + \omega_{MN} \right) {E_{[A}}^M(\sigma_1) \partial_I {E_{B]}}^N(\sigma_1) \delta(\sigma_3 - \sigma_1) \delta(\sigma_2 - \sigma_1). \nonumber
\end{equation}
As a cross check we obtain the same form of $\mathbf{Q}$-flux non-associativity in the first line of \eqref{eq:JacobiIdNoSC} by inserting the generalised vielbein to the $\mathbf{Q}$-flux background and integrating accordingly as before. All the other terms in \eqref{eq:JacobiIdNoSC} vanish in this simple example.

\paragraph{The other $T$-duality chain backgrounds}
The non-associativity will not be relevant if we only 'probe' the phase space with functions $f(y^a;\bE_A)$ resp. $f(x^a;\bE_I)$. As $\tilde{y}_3$ is not an argument of these functions the $\mathbf{Q}$-flux background given by the current algebra \eqref{eq:QFluxCurrentAlgebra} is \textit{associative} and thus locally geometric. But there are other different choices of solutions of the strong constraints\footnote{We phrase them in the phase space variables of the $\mathbf{Q}$-flux background. To get the standard picture, e.g. of the $\mathbf{H}$-flux we make the identifications $y^1 \leftrightarrow \tilde{y}_1$ and $y^2 \leftrightarrow \tilde{y}_2$.} which correspond to the $T$-dual backgrounds of the $T$-duality chain (see section \ref{chap:reviewGeneralisedFluxes}):
\begin{align*}
f(y^1,y^2,y^3;...) \quad &{} \quad \text{locally geometric $\mathbf{Q}$-flux background}, \\
f(\tilde{y}_1,y^2,y^3;...) \ \text{or} \ f(y^1,\tilde{y}_2,y^3;...) \quad &{} \quad \text{locally geometric $\mathbf{f}$-flux backgrounds}, \\
f(\tilde{y}_1,\tilde{y}_2,y^3;...) \quad &{} \quad \text{locally geometric $\mathbf{H}$-flux background}.
\end{align*}
In addition there are of course also the continuous O$(2,2)$-transformations on the $y_1,y_2$.

The solutions of the strong constraint containing $\tilde{y}_3$ give non-associative phase spaces, corresponding to the locally non-geometric backgrounds:
\begin{align*}
f(y^1,y^2,\tilde{y}_3;...) \quad &{} \quad \text{locally non-geometric $\mathbf{R}$-flux background}, \\
f(\tilde{y}_1,y^2,\tilde{y}_3;...) \ \text{or} \ f(y^1,\tilde{y}_2,\tilde{y}_3;...) \quad &{} \quad \text{locally non-geometric $\mathbf{Q}$-flux backgrounds}, \\
f(\tilde{y}_1,\tilde{y}_2,\tilde{y}_3;...) \quad &{} \quad \text{locally non-geometric $\mathbf{f}$-flux background}.
\end{align*}
These are all locally non-geometric as the generalised vielbein depends via $\beta$ on $x^3 = y^3$, which is the origin of the non-associativity.

Overall we reproduce the well-known zero mode brackets and non-vanishing Jacobiators \cite{Blumenhagen:2010hj,Lust:2010iy,Blumenhagen:2011ph,Condeescu:2012sp,Andriot:2012an,Chatzistavrakidis:2012qj,Mylonas:2012pg,Bakas:2013jwa,Plauschinn:2018wbo} of the considered (non-geometric) backgrounds without imposing a mode expansion or the equations of motion.

\section{Discussion}

\subsection{Summary}

The central result of this paper was introduced in section \ref{chap:Hamiltonian}.1. The world-sheet theory in a generic NSNS background, including non-geometric ones, can be defined in the following way. In terms of some phase space variables $\bE_A(\sigma)$ there is a Hamiltonian in a background independent form $H \sim \int \mathrm{d}\sigma \ \delta^{AB} \bE_A(\sigma) \bE_B(\sigma)$, and similarly for the Virasoro constraints. Instead the information about the background is encoded in the Poisson structure. This is most conveniently formulated in terms of the current algebra (the algebra of the $\bE_A(\sigma)$)
\begin{equation}
\PB{\bE_A(\sigma_1)}{\bE_B(\sigma_2)} = \Pi^{\eta}_{AB}(\sigma_1,\sigma_2) + \Pi^{\text{bdy.}}_{AB}(\sigma_1,\sigma_2) + \Pi^{\text{flux}}_{AB}(\sigma_1,\sigma_2) \label{SUMMARY-PB}
\end{equation}
$\Pi^{\eta}$ is the O$(d,d)$-invariant part of the canonical current algebra \eqref{eq:PBCanonicalCurrent}, whereas
\begin{align*}
\Pi^{\text{flux}}_{AB}(\sigma_1,\sigma_2) = - \bF_{ABC}(\sigma_1) \bE^C(\sigma_1) \delta(\sigma_2 - \sigma_1)
\end{align*}
is characterised solely by the generalised flux $\bF_{ABC}$, building on known results in the literature \cite{Alekseev:2004np,Halmagyi:2008dr,Halmagyi:2009te}. This formulation seems to be the world-sheet version of the generalised flux formulation of generalised geometry resp. double field theory \cite{Grana:2008yw,Geissbuhler:2013uka}.

In case of an electric and locally geometric background, meaning the Bianchi identity \eqref{eq:BianchiIdGeneralisedFluxes} is fulfilled, there is a connection to Darboux coordinates $(x^i,p_i)$ on the phase space resp. a Lagrangian formulation. This connection is given by a choice of generalised vielbein ${E_A}^I(c)$, s.t. $\bE_A(\sigma) = {E_A}^I(x(\sigma)) (p_i(\sigma) , \partial x^i(s\sigma))$ and $\bF_{ABC} = (\partial_{[A} {E_B}^I) E_{C]I}$. 

In the cases of a magnetically charged NSNS background (like an NS5-brane) or a locally non-geometric background the Hamiltonian world-sheet theory as define above is still defined. But there are some obstructions in either case. In former the Bianchi identity of generalised fluxes is sourced. Resultantly the associated current algebra violates the Jacobi identity and thus there cannot be Darboux coordinates on the phase space associated to the sourcing world-volume. In the non-geometric case, see section \ref{chap:DFT} and \ref{chap:NonGeometry} for more details, there will be a violation of the Jacobi identity if we consider certain functions of the 'doubled' string phase space. E.g. for the Jacobi identity of a functional $\Psi$ and two sections $\phi_i = \oint \mathrm{d} \sigma \phi^I_i(\sigma) \bE_I(\sigma)$ we obtained
\begin{align}
\PB{\Psi}{\PB{\phi_1}{\phi_2}} + c.p. &= \oint \mathrm{d} \sigma_1 \mathrm{d}\sigma_2 \frac{1}{2} (\eta_{JK} + \omega_{JK}) \phi_{[1}^J(\sigma_1) \frac{\delta \Psi}{\delta X^I(\sigma_2)} \frac{\delta \phi_{2]}^K(\sigma_1)}{\delta X_I (\sigma_2)}  \label{SUMMARY:JacobiIdNoSC} + \ \text{others.}
\end{align}
In case the generalised vielbein itself depends an original coordinate and its dual at the same time a additional term in \eqref{SUMMARY-PB} appears that potentially leads to a non-local contribution to the equations of motion and a modication of the Virasoro algebra. 

One difference to previous discussions in the literature is the consideration of the total derivative term,
\begin{align*}
\Pi^{\text{bdy.}}_{AB}(\sigma_1,\sigma_2) = \int \mathrm{d}\sigma \partial \left( \omega_{AB} (\sigma) \delta(\sigma - \sigma_1) \delta(\sigma - \sigma_2) \right).
\end{align*}
This occurs in this form as a \textit{non} O$(d,d)$-invariant boundary contribution from the canonical current algebra \eqref{eq:PBCanonicalCurrent}. Terms like this in the current algebra itself or its Jacobi identity make the difference between a Lie or a Courant algebroid structure of the phase space $(T \oplus T^\star) LM$. This was discussed in detail in section \ref{chap:CourantAlgebroid}. For open strings they lead to the known constraint of $\mathbf{H}\big\vert_\text{D-brane}=0$ \cite{Alekseev:2004np} and the non-commutativity at the ends of the open string \cite{Chu:1998qz}. For closed strings a winding contribution from this term is necessary such that the standard $\mathbf{Q}$-flux background is an associative background.

We discussed two applications of this formulation of the world-sheet theory. The first one is the observation that (generalised) $T$-dualities act linearly on the variables in the generalised flux frame. This lead to the proposal of a generalisation of Poisson-Lie $T$-duality to \textit{Roytenberg duality}, applicable to models with constant generalised fluxes. This was shown using a certain parameterisation of the constant generalised flux based on the ones of Poisson-Lie $\sigma$-models in section \ref{chap:Hamiltonian}.2.

The second application is a direct derivation of the well-known non-commutative and non-associative behaviour of some generalised flux backgrounds from the deformed current algebra in section \ref{chap:NonGeometry}. This interpretation does not rely on a mode expansion or even on imposing the equations of motion, it is purely kinematic. Also it extends straightforwardly to any generalised flux background.

\subsection{Potential applications and open problems}

Part of the original motivation was the study of integrable deformation, as these can be conveniently represented as deformations of the current algebra -- see section \ref{chap:IntModel}. The discussion in this paper connecting the possible deformations of the current algebra for string $\sigma$-models to generalised fluxes, hints at a connection of generalised geometry to the Hamilton formulation of integrable $\sigma$-models. From a purely technical side there is also a argument to maybe expect a connection to integrability. The currents $\mathbf{e}_a$, used here to write down the equations of motion \eqref{eq:GeneralisedFluxEOM}, are the ones which are used to calculate the Lax pair in all the examples -- principal chiral model, $\eta$-deformation, $\lambda$-deformation, Yang-Baxter deformation.

There are two generalisations of this article's approach which come to mind immediately. The first one is the generalisation to the \textit{Green-Schwarz superstring}, whereas an RNS formulation was already given in \cite{Blair:2013noa} in context of the generalised metric formulation. In particular introducing $RR$-fluxes into the deformation of the current algebra could be interesting to obtain a direct understand the world-sheet dynamics in an $RR$-flux backgrounds -- as generically the $RR$-flux terms in the non-linear $\sigma$-model are not known explicitly. For the Green-Schwarz superstring a complete kinematic description includes $\kappa$-symmetry, which on the other hand is also closely connected to the supergravity equations \cite{Wulff:2016tju} and thus dynamics of the background. The fact that this formalism relies on a flat internal space might be useful to define spacetime fermions in a background independent way and a formulation of the Green-Schwarz superstring, that is not only valid in very symmetric spacetimes. In principle the generalised flux formulation of the current algebra in the Green-Schwarz approach was given already in \cite{Siegel:1993th}, but without the topological term and without a non-geometric interpretation of the superversion of the generalised fluxes $\bF_{ABC}$ occuring there. 

Another generalisation would be to the Hamiltonian treatment of \textit{membrane $\sigma$-models}. There has been a lot of work on topological membrane $\sigma$-models. An approach similar to the one discussed here could be useful to understand the kinetic term of membrane $\sigma$-models better. Also the appearance and the interpretation of expected higher brackets in the phase space of a membrane seems interesting to study. Very recently a generalisation of Poisson-Lie $T$-duality to higher gauge theories was proposed \cite{Pulmann:2019vrw}, it would be interesting to investigate whether such dualities are realised in a membrane current algebra in a similarly simple fashion as (generalised) $T$-dualities here.

As demonstrated in \cite{Halmagyi:2009te} it is not advantageous to parameterise the background by the generalised fluxes in order to calculate the 1-loop $\beta$-function and check the quantum conformality like this. But this formulation might be potentially a good framework to quantise the string canonically. In particular for constant generalised fluxes the equations of motion \eqref{eq:GeneralisedFluxEOMCov} take the form of a (constrained) Maurer-Cartan structure equations of a $2d$-dimensional (non-compact) Lie group. If it would be possible to construct a mode expansion, it seems possible to quantise the bosonic theory directly as also the Virasoro constraint take a simple form in the generalised flux frame.

An open technical problem is the relation of the canonical (deformed) current algebra \eqref{SUMMARY-PB} including topological/total derivative terms and the 'double field theory' algebra $\PB{X_I(\sigma)}{X_J(\sigma^\prime)} = - \eta_{IJ} \bar{\Theta}(\sigma - \sigma^\prime)$, as they are not equivalent. It would be very useful to understand this better as in this relation seems to lie the source of the non-associativity associated to strong constrain violations in section \ref{chap:DFT}. Also, it was mentioned before that apart from the fact that we assumed our generalised fluxes to be globally well-defined tensors we only discussed local properties of our globally non-geometric backgrounds. Previous work discussing current algebras, loop algebras and their global properties is \cite{Belov:2007qj, Hekmati:2012fb}. Connecting these approaches and the generalised flux formulation of non-geometric background seems to be an important step for future work.

\subsection*{Acknowledgements}
I thank Falk Hassler, Dieter L\"ust and Matthias Traube for useful discussions, Henk Bart, Dieter L\"ust and Andriana Makridou for comments on the draft, and especially Chris Blair and Warren Siegel for helpful remarks on the first version.

\appendix

\thispagestyle{empty}
\bibliographystyle{jhep}
\bibliography{References}

\end{document}